\documentclass[11pt]{article}

\usepackage{indentfirst}
\usepackage{amsmath}
\usepackage{amssymb}
\usepackage{amsfonts}
\usepackage[dvipsnames]{xcolor}
\usepackage[pdftex]{graphicx}
\usepackage{subfigure}
\usepackage{multirow}
\usepackage{makecell}
\usepackage{float}

\usepackage{geometry}
\usepackage[backend=bibtex, style=numeric-comp, sorting=none, maxnames=99]{biblatex}
\usepackage{hyperref}
\hypersetup
{
    colorlinks = true,
    linkcolor = red,
    citecolor = blue,
    urlcolor = PineGreen
}

\graphicspath{{./Figs/}}

\begin{document}

\begin{titlepage} \setcounter{page}{0}
\begin{center}
    \vspace*{1.0cm}
    {\Large\bf Frame Dependence of Bound on Lyapunov Exponent in Dilatonic Reissner--Nordstr\"om--AdS and Kerr--Sen--AdS Black Holes}
    \\ \vspace{2.0cm}
    {$\mbox{Hocheol Lee}$}\footnote{\it email: insaying@dongguk.edu}, \quad
    {$\mbox{Bogeun Gwak}$}\footnote{\it email: rasenis@dgu.ac.kr}
    \\ \vspace{0.2cm}
    {\small \it Department of Physics, Dongguk University, Seoul 04620, Republic of Korea}
    \\ \vspace{2.0cm}
\end{center}

\begin{center}
\begin{abstract}
    We investigate the frame dependence of the Lyapunov exponent bound for charged particles in dilatonic Reissner--Nordstr\"om--AdS and Kerr--Sen--AdS black hole backgrounds, derived from Einstein--Maxwell--dilaton theory and the low-energy effective action of heterotic string theory, respectively. The analysis is performed in both the Einstein and string (Jordan) frames to examine the influence of conformal transformations on chaotic behavior. For massless particles, the Lyapunov exponent remains invariant under frame transformations, whereas for massive particles, it exhibits frame dependence owing to coupling to the dilaton field. Our results indicate sensitivity of the bound on chaos to the choice of frame. Depending on various parameters, the bound can be satisfied in the Einstein frame and violated in the string frame, while the opposite situation may occur for different parameter values. Numerical computations corroborate the findings of our analysis and demonstrate modifications in the chaotic behavior of string-inspired black holes induced by the dilaton field and the choice of frame.
\end{abstract}
\end{center}
\end{titlepage}

\section{Introduction} \label{sec_intro}
    Black holes, initially recognized as solutions of general relativity whose singularities and event horizons challenged conventional intuition~\cite{Schwarzschild:1916uq}, have evolved into fundamental objects in modern theoretical physics, offering profound insights into the nature of spacetime, gravity, and quantum mechanics. Beyond geometric characteristics, the concepts and open questions exemplified by multiple horizons~\cite{Reissner:1916cle, Kerr:1963ud}, black hole entropy~\cite{Bekenstein:1973ur}, Hawking radiation~\cite{Hawking:1975vcx}, information loss~\cite{Hawking:1976ra, Page:1993wv}, primordial black holes~\cite{Zeldovich:1967lct, Hawking:1971ei}, and wormholes~\cite{Einstein:1935tc, Morris:1988cz} highlight further challenges to understanding classical and quantum gravity. These unresolved problems render black holes particularly compelling subjects for exploring deeper theoretical constructs, notably the anti--de Sitter\slash conformal field theory (AdS\slash CFT) correspondence~\cite{Maldacena:1997re}, which relates gravitational dynamics in the AdS bulk to the conformal field theory defined on its boundary. The study of black holes in AdS spacetimes constitutes a foundational element in the development of holography~\cite{tHooft:1993dmi, Susskind:1994vu, Witten:1998qj}, providing insights into the connection between gravitational theories and boundary conformal field theories. Investigations of AdS black holes have revealed a wide range of structural and dynamical properties, enriching the understanding of gravitational systems and the corresponding holographic duals, and offering a versatile approach to analyzing strongly coupled quantum theories. Research on black holes elucidates essential aspects of quantum gravity, including the interplay between geometry, thermodynamics, and quantum information, establishing a theoretical structure for probing the fundamental principles of physics.

    Higher-dimensional~\cite{Emparan:2008eg, Horowitz:2012nnc} and string-inspired black holes~\cite{Callan:1988hs, Mignemi:1992nt} have been extensively studied in theoretical physics, particularly in the context of the AdS\slash CFT correspondence~\cite{Witten:1998zw, Hawking:1998kw, Chamblin:1999tk}. String-inspired models extend classical general relativity by introducing additional fields and couplings, including the dilaton, antisymmetric tensor fields, gauge fields, and higher-curvature terms~\cite{Callan:1985ia, Zwiebach:1985uq, Gross:1986iv}. The dilaton is a scalar field arising naturally from string theory, coupling nonminimally to the metric and functioning as an overall factor, resulting in two widely used formulations: the Einstein frame and the string (Jordan) frame. The two frames are related by a conformal transformation, which corresponds to a rescaling of the metric and a modification of the couplings of the fields. In the Einstein frame, gravitational action is expressed in canonical form, enabling direct comparison with general relativity and the application of standard techniques in classical gravitational analysis. By contrast, the string frame naturally emerges when describing the low-energy effective theory of string theory. The local conformal invariance of the worldsheet theory requires the beta functions of all background fields to vanish. The vanishing of beta functions eliminates trace anomalies and imposes constraints on the target space fields. These constraints can be derived from an effective spacetime action formulated in the string frame, which in higher-dimensional string theory can be further reduced to an effective four-dimensional action through dimensional reduction or compactification~\cite{Witten:1985xb, Duff:1994tn}. The four-dimensional effective action introduces additional dynamic degrees of freedom and modifies gravitational dynamics, reflecting the structure of the underlying higher-dimensional theory, with the string frame providing the original form of the effective action. In contrast, the Einstein frame is conventionally employed because it expresses the gravitational action in canonical form and facilitates the application of standard techniques in black hole research.

    The Einstein and string frames are formally connected through conformal transformations. Observables and physical quantities, including the formulation itself, can exhibit frame-dependent features, suggesting that the representation of the underlying physics may vary between frames. In certain contexts, specific frames offer more consistent descriptions based on theoretical or phenomenological criteria. Studies on frame dependence have been extensively examined and reviewed in~\cite{Magnano:1993bd, Faraoni:1998qx}. These studies can be categorized as follows: \textbf{(\romannumeral 1)}~studies acknowledging the potential physical non-equivalence of conformally related frames without advocating a preferred frame~\cite{Capozziello:2010sc, Domenech:2015qoa, Banerjee:2016lco, Karam:2017zno, Chakraborty:2023kel}, \textbf{(\romannumeral 2)}~those supporting the physical equivalence of theories in different frames~\cite{Steinwachs:2013tr, Chiba:2013mha, Postma:2014vaa, Kamenshchik:2014waa, Burns:2016ric, Jarv:2016sow, GiontiSJ:2023tgx}, \textbf{(\romannumeral 3)}~those identifying the Einstein frame as physical (or conventionally adopting it as the canonical frame), and \textbf{(\romannumeral 4)}~those identifying the string frame as physical~\cite{Sakstein:2015jca, Ko:2016dxa, Rinaldi:2018qpu, Lee:2023boi}. Currently, research studies are continuing to explore these questions to further clarify the role of conformal frames in theoretical and phenomenological contexts.

    Past research studies have reported frame dependence in various physical observables. The present work extends the line of inquiry to the Lyapunov exponent. The classical Lyapunov exponent quantifies the sensitivity of a dynamical system to variations in initial conditions. In the context of the AdS\slash CFT correspondence, the Lyapunov exponent is regarded as a key diagnostic of quantum chaos. Maldacena, Shenker, and Stanford~\cite{Maldacena:2015waa} demonstrated that after the dissipation time, the out-of-time-order correlator in a thermal quantum system grows exponentially at a rate $\lambda$, analogous to the classical Lyapunov exponent, and conjectured a universal bound $\lambda \leq 2 \pi T$, where $T$ denotes the system temperature. This result suggests a potential link between gravitational dynamics and quantum chaos by identifying the Lyapunov exponent as a quantity that connects classical instability with quantum information scrambling. In a subsequent study, Hashimoto and Tanahashi~\cite{Hashimoto:2016dfz} proposed a correspondence between the temperature $T$ of a quantum system and the Hawking temperature $T_{\mathrm{H}}$ of a black hole. The analysis of particle orbits near the black hole horizon revealed that the Lyapunov exponent, interpreted as a measure of orbital instability in the presence of additional fields, such as a Maxwell field or a massless scalar, remains consistent with the bound on chaos $\lambda \leq 2 \pi T_{\mathrm{H}}$. The results demonstrate the applicability of the bound on chaos in gravitational systems, exhibiting classical orbital dynamics with characteristics consistent with universal features of quantum chaotic behavior. Lyapunov exponents have been investigated across a wide range of black hole backgrounds~\cite{Kan:2021blg, Gwak:2022xje, Yu:2022tlr, Gao:2022ybw, Chen:2022tbb, Hashimoto:2022kfv, Chakrabortty:2022kvq, Yin:2022mjv, Song:2022lhf, Jeong:2023hom, Chen:2023wph, Yu:2023spr, Li:2023bgn, Lei:2023jqv, Prihadi:2023tvr, Prihadi:2023qmk, Xie:2023tjc, Dutta:2023yhx, Karan:2023hfk, Park:2023lfc, Han:2023ckr, Kumara:2024obd, Giataganas:2024hil, Lei:2024qpu, Das:2024iuf, Singh:2024qfw, Dutta:2024rta, Gallo:2024wju, Ciou:2025ygb, R:2025gok, Lee:2025vih} and, more recently, have also been proposed as probes of black hole phase transitions~\cite{Guo:2022kio, Yang:2023hci, Lyu:2023sih, Du:2024uhd, Shukla:2024tkw, Gogoi:2024akv, Awal:2025irl, Yang:2025fvm, Kumar:2025kzt, Guo:2025pit, Bezboruah:2025udi, Zhang:2025cdx}. However, although research studies thus far have extensively investigated Lyapunov exponents, to the best of our knowledge the potential dependence on the choice of conformal frame has not yet been systematically addressed. Because conformal transformations can modify the representation of physical quantities, examining frame dependence provides a critical test of the robustness of chaos signatures under different theoretical formulations.

    To investigate the Lyapunov exponent in gravitational systems, the present analysis considers the dilatonic Reissner--Nordstr\"om (RN) black hole and the Kerr--Sen black hole as representative cases. The dilatonic RN black hole is a solution of the Einstein--Maxwell--dilaton (EMD) theory. Gibbons and Maeda~\cite{Gibbons:1987ps} obtained a static, spherically symmetric, charged black hole solution of the EMD theory that includes a 2-form field and a $(D-2)$-form field, where $D$ denotes the dimension of spacetime. Independently, Garfinkle, Horowitz, and Strominger~\cite{Garfinkle:1990qj} derived an analogous solution from the low-energy effective action of heterotic string theory that is restricted to the magnetic charge and omits the antisymmetric three-form field. The resulting solution is commonly referred to as the Gibbons--Maeda--Garfinkle--Horowitz--Strominger (GMGHS) black hole. The Kerr--Sen black hole extends the GMGHS solution by introducing rotation. Sen~\cite{Sen:1992ua} applied the Hassan--Sen transformation~\cite{Hassan:1991mq} to the Kerr metric within the low-energy effective action of heterotic string theory, from which the GMGHS solution is derived, producing a charged, rotating solution incorporating the dilaton and the antisymmetric two-form field. The Kerr--Sen black hole can be understood as a rotating generalization of the GMGHS black hole, allowing for the study of how string-inspired effects manifest in both static and rotating black hole spacetimes.

    In the present work, the Lyapunov exponent is analyzed in the dilatonic RN--AdS and Kerr--Sen--AdS backgrounds, with the analyses performed in both the Einstein and string frames. Classical black hole solutions, including Kerr--Newman~\cite{Kan:2021blg}, Kerr--Newman--AdS~\cite{Gwak:2022xje}, and Kerr--Newman--dS~\cite{Park:2023lfc} spacetimes, have been studied in past research efforts. The Kerr--Sen--AdS black hole has been examined in the Einstein frame~\cite{Lee:2025vih} to assess modifications arising from string-inspired theory. In the present study, both the dilatonic RN--AdS and Kerr--Sen--AdS black holes are analyzed in the Einstein and string frames. For each black hole, the effective potential for particle motion is derived, and the Lyapunov exponent is computed. The results demonstrate that for massless particles, the Lyapunov exponent remains identical between frames, whereas for massive particles, the exponent exhibits differences. The analysis examines the influence of string-inspired modifications on the bound on chaos, which would indicate frame-dependent variations in chaotic behavior.

    The remainder of this paper is organized as follows. Section~\ref{sec_review} reviews the formulation of the Lyapunov exponent. Section~\ref{sec_frame_lyapunov_exponent} provides an analysis of the Lyapunov exponent in different frames and compares its values. Section~\ref{sec_RN} presents an investigation of the Lyapunov exponent and the bound on chaos in the dilatonic RN--AdS black hole, considering both the effects of the coupling constant and its dependence on the choice of frame. Section~\ref{sec_KS} examines the Lyapunov exponent and the bound on chaos in the Kerr--Sen--AdS black hole, addressing the influence of relevant parameters and frame dependence. Finally, Section~\ref{sec_conclusion} summarizes the main results and discusses possible directions for future research.

\section{Review of Lyapunov Exponent} \label{sec_review}
    The Lyapunov exponent quantifies the sensitivity of classical particle trajectories to initial conditions, capturing the exponential divergence of nearby paths and functioning as an indicator of chaotic behavior. To analyze the dynamical stability systematically, we introduce a generalized Lagrangian for radial motion, expressed as
\begin{equation}
    \mathcal{L}(r, \dot{r}) = \frac{1}{2} K(r) \dot{r}^2 - V(r),
    \label{eq:review_lagrangian}
\end{equation}
\noindent where $K(r)$ is the radial-dependent kinetic term reflecting the background geometry, and the dot denotes the derivatives with respect to the geodesic parameter. The radial momentum is obtained from the Lagrangian~\eqref{eq:review_lagrangian} as
\begin{equation}
    p_r \equiv \frac{\partial \mathcal{L}}{\partial \dot{r}} = K \dot{r}.
\end{equation}
    A Legendre transformation is performed, and the Hamiltonian $H$ is obtained as follows:
\begin{equation}
    H(r, p_r) \equiv p_r \dot{r} - \mathcal{L} = \frac{p_r^2}{2 K} + V.
    \label{eq:hamiltonian}
\end{equation}
    From the Hamiltonian~\eqref{eq:hamiltonian}, the equations of motion are given by
\begin{equation}
    \dot{r} = \frac{\partial H}{\partial p_r}, \qquad \dot{p}_r = - \frac{\partial H}{\partial r},
\end{equation}
\noindent which can be combined in matrix form as
\begin{equation}
    \frac{\mathrm{d}}{\mathrm{d}\tau}
    \begin{pmatrix} 
        r
        \\[1.0mm]
        p
    \end{pmatrix}
    =
    \begin{pmatrix} 
        0 & 1
        \\[1.5mm]
        -1 & 0
    \end{pmatrix}
    \begin{pmatrix} 
        \frac{\partial H}{\partial r}
        \\[1.5mm]
        \frac{\partial H}{\partial p_r}
    \end{pmatrix}
    =
    \begin{pmatrix} 
        \frac{p_r}{K}
        \\[1.5mm]
        \frac{1}{2} K' \dot{r}^2 - V'
    \end{pmatrix},
    \label{eq:review_hamilton_equations}
\end{equation}
\noindent where the prime symbol indicates the derivative with respect to the radial coordinate $r$. At the radial equilibrium point, $r_0$, the radial velocity and momentum vanish:
\begin{equation}
    \left. \dot{r} \right|_{r = r_0} = \left. p \right|_{r = r_0} = 0,
\end{equation}
\noindent and the potential satisfies
\begin{equation}
    V'(r_0) = 0.
\end{equation}
    To quantify the local instability of the orbit around $r_0$, we introduce small deviations $\delta r$ and $\delta p_r$:
\begin{equation}
    r = r_0 + \delta r, \qquad p = \delta p.
    \label{eq:review_deviation}
\end{equation}
    Substituting~\eqref{eq:review_deviation} into Eq.~\eqref{eq:review_hamilton_equations} and expanding around $r_0$, we keep solely the leading-order terms in $\delta r$ and $\delta p_r$\,. The resulting linearized equations of motion are
\begin{equation}
    \frac{\mathrm{d}}{\mathrm{d}\tau}
    \begin{pmatrix} 
        \delta r
        \\[1.5mm]
        \delta p
    \end{pmatrix}
    =
    \begin{pmatrix} 
        0 & \frac{1}{K(r_0)}
        \\[1.5mm]
        - V''(r_0) & 0
    \end{pmatrix}
    \begin{pmatrix} 
        \delta r
        \\[1.0mm]
        \delta p
    \end{pmatrix}.
    \label{eq:review_linearized_hamilton_equations}
\end{equation}
    The local Lyapunov exponent $\lambda$ is given by the eigenvalues of the linearized matrix of Eq.~\eqref{eq:review_linearized_hamilton_equations}:
\begin{equation}
    \lambda^2 = - \frac{V''(r_0)}{K(r_0)}.
    \label{eq:reveiw_lyapunov_exponent}
\end{equation}
    For an unstable orbit, where $V'' < 0$, the Lyapunov exponent~\eqref{eq:reveiw_lyapunov_exponent} takes a positive value. The positive Lyapunov exponent quantifies the exponential divergence of nearby trajectories and provides a measure of the local chaotic behavior of the system.

\section{Frame Dependence of Lyapunov Exponent} \label{sec_frame_lyapunov_exponent}
    We study the dynamics of a particle to analyze the frame dependence of the Lyapunov exponent. The Polyakov-type Lagrangian for an electrically charged particle is given by
\begin{equation}
    \mathcal{L}_\mathrm{P} = \frac{1}{2 \sigma(s)} \left( \frac{\mathrm{d}\mathbf{X}}{\mathrm{d}s} \right)^2 - \frac{\sigma(s)}{2} m^2 + q A_\mu \frac{\mathrm{d}X^\mu}{\mathrm{d}s},
\end{equation}
\noindent where $m$ and $q$ are the mass and electric charge of the particle, respectively; $\sigma$ is an auxiliary field; s parametrizes the geodesic of the particle; $A_\mu$ is the electromagnetic potential; and $X^\mu$ denotes the spacetime coordinates, where $\mathbf{X} = \left\{ t(s), r(s), \theta(s), \varphi(s) \right\}$. We adopt the stationary, axisymmetric, and circular metric in the string frame, defined by
\begin{eqnarray}
    \mathrm{d}s_\mathrm{S}^2 &\equiv& e^{\phi} \mathrm{d}s_\mathrm{E}^2
    \\
    &=& e^{\phi(r, \theta)} \left[ g_{tt}(r, \theta)  \mathrm{d}t^2 + 2 g_{t\varphi}(r, \theta) \mathrm{d}t \mathrm{d}\varphi + g_{rr}(r, \theta) \mathrm{d}r^2 + g_{\theta \theta}(r, \theta) \mathrm{d}\theta^2 + g_{\varphi \varphi}(r, \theta) \mathrm{d}\varphi^2 \right],
    \label{eq:metric}
\end{eqnarray}
\noindent where S denotes the string frame, E signifies the Einstein frame, $g_{\mu\nu}$ represents the metric tensor in the Einstein frame, and $\phi$ represents the dilaton field. The electromagnetic potential associated with the aforementioned metric~\eqref{eq:metric} takes the form
\begin{equation}
    \mathbf{A} = A_t(r, \theta) \mathrm{d}t + A_\varphi(r, \theta) \mathrm{d}\varphi.
\end{equation}
    We choose the static gauge $t = s$ and restrict the motion to the equatorial plane, {\it i.e.}, $\theta = \pi/2$. Given the static gauge and the equatorial plane restriction, the Lagrangian for a particle in the string frame is reduced to
\begin{equation}
    \mathcal{L}_\mathrm{P} = \left. \frac{e^{\phi}}{2 \sigma} \left( g_{tt} + g_{rr} \dot{r}^2 + g_{\varphi\varphi} \dot{\varphi}^2 + 2 g_{t\varphi} \dot{\varphi} \right) - \frac{\sigma}{2} m^2 + q \left( A_t + A_\varphi \dot{\varphi} \right) \right|_{\theta = \frac{\pi}{2}}.
    \label{eq:lagrangian_particle}
\end{equation}
    Because the aforementioned Lagrangian~\eqref{eq:lagrangian_particle} is independent of $\varphi$, the angular momentum of a particle in the string frame, as a conserved quantity, is given by
\begin{equation}
    L \equiv \frac{\partial \mathcal{L}_\mathcal{P}}{\partial \dot{\varphi}} = \frac{e^\phi}{\sigma} \left( g_{\varphi\varphi} \dot{\varphi} + g_{t\varphi} \right) + q A_\varphi.
\end{equation}
    From the equation of motion for the auxiliary field $\sigma$, the constraint $\dot{X}^2 = - \sigma^2 m^2$ follows, and $\sigma$ is determined to be
\begin{equation}
    \sigma^2 = \frac{e^{2 \phi} \left[ g_{t\varphi}^2 - g_{\varphi\varphi} \left( g_{tt} + g_{rr} \dot{r}^2 \right) \right]}{\left( L - q A_\varphi \right)^2 + m^2 e^\phi g_{\varphi\varphi}}.
\end{equation}
    Substituting the expressions for the conserved angular momentum $L$ and auxiliary field $\sigma$ yields the effective Lagrangian $\mathcal{L}_\mathrm{eff}$, defined as
\begin{eqnarray}
    \mathcal{L}_\mathrm{eff}(r, \dot{r}) \!\!\!\! &\equiv& \!\!\!\! \mathcal{L}_\mathrm{P} - L \dot{\varphi}
    \\
    &=& \!\!\!\! \left. \frac{1}{g_{\varphi\varphi}} \! \left[ \left( L \! - \! q A_\varphi \right) \! g_{t\varphi} \! + \! q A_t g_{\varphi\varphi} \! - \! \sqrt{\left( L \! - \! q A_\varphi \right)^2 \! + \! m^2 e^\phi g_{\varphi \varphi}} \sqrt{g_{t\varphi}^2 \! - \! g_{\varphi\varphi} \! \left( g_{tt} \! + \! g_{rr} \dot{r}^2 \right)} \right] \right|_{\theta = \frac{\pi}{2}}.
    \label{eq:original_effective_lagrangian}
\end{eqnarray}
    The dynamics are simplified to a one-dimensional system in the radial coordinate $r$ by defining the effective Lagrangian. We investigate the chaotic behavior of the particle in the non-relativistic limit $(\dot{r} \ll 1)$, where the effective Lagrangian~\eqref{eq:original_effective_lagrangian} reduces to
\begin{equation}
    \mathcal{L}_\mathrm{eff} = \frac{1}{2} K_\mathrm{S}(r) \dot{r}^2 - V_\mathrm{S}(r) + \mathcal{O}(\dot{r}^4),
    \label{eq:nonrelativistic_effective_lagrangian}
\end{equation}
    where
\begin{eqnarray}
    K_\mathrm{S}(r) \!\! &=& \!\! \left. g_{rr} \sqrt{\frac{\left( L - q A_\varphi \right)^2 + m^2 e^\phi g_{\varphi \varphi}}{g_{t\varphi}^2 - g_{tt} g_{\varphi\varphi}}} \right|_{\theta = \frac{\pi}{2}},
    \label{eq:string_kinetic_term}
    \\
    V_\mathrm{S}(r) \!\! &=& \!\! \left. \frac{1}{g_{\varphi \varphi}} \sqrt{\left( L - q A_\varphi \right)^2 + m^2 e^\phi g_{\varphi \varphi}} \sqrt{g_{t\varphi}^2 - g_{tt} g_{\varphi\varphi}} - L \frac{g_{t\varphi}}{g_{\varphi\varphi}} - q \left( A_t - \frac{g_{t\varphi}}{g_{\varphi\varphi}} A_\varphi \right) \right|_{\theta = \frac{\pi}{2}}.
    \label{eq:string_effective_potential}
\end{eqnarray}
    Focusing on the behavior near the local maximum of the effective potential, we consider the particle initially at rest at $r_0$, where $V'(r_0) = 0$ and $V''(r_0) < 0$, and introduce a small deviation $\epsilon$ around $r_0$\,. With the small deviation $\epsilon$, the effective Lagrangian~\eqref{eq:nonrelativistic_effective_lagrangian} can be expanded into
\begin{equation}
    \mathcal{L}_\mathrm{eff} = \frac{1}{2} K_\mathrm{S}(r_0) \left( \dot{\epsilon}^2 + \lambda^2 \epsilon^2 \right),
    \label{eq:effective_lagrangian}
\end{equation}
\noindent where the constant term has been omitted. The Lyapunov exponent in string frame $\lambda_\mathrm{S}$ is obtained from the coefficient of the quadratic term in the effective Lagrangian~\eqref{eq:effective_lagrangian} according to
\begin{equation}
    \lambda_\mathrm{S}^2 = - \frac{V_\mathrm{S}''(r_0)}{K_\mathrm{S}(r_0)}.
    \label{eq:string_Lyapunov_exponent}
\end{equation}
    The result coincides with the Lyapunov exponent in~\eqref{eq:reveiw_lyapunov_exponent}. The analysis presented in this section has been performed in the string frame. In general, the Lyapunov exponent in the string frame~\eqref{eq:string_Lyapunov_exponent} differs from the Lyapunov exponent in the Einstein frame:
\begin{equation}
    \lambda_\mathrm{E}^2 \neq \lambda_\mathrm{S}^2.
\end{equation}
    The Lyapunov exponent in the Einstein frame can be obtained from the string frame expression by imposing the condition $\phi = 0$, which yields
\begin{equation}
    \lambda_\mathrm{E}^2 = \left. \lambda_\mathrm{S}^2 \right|_{\phi = 0}.
\end{equation}
    The frame dependence originates from the coupling between the dilaton and the metric. The explicit forms of the kinetic term~\eqref{eq:string_kinetic_term} and the effective potential~\eqref{eq:string_effective_potential} show that the particle mass $m$ effectively represents the contribution of the dilaton. In the massless limit, the kinetic term~\eqref{eq:string_kinetic_term} and the effective potential~\eqref{eq:string_effective_potential} are identical in the Einstein and string frames, with the corresponding expressions being reduced to
\begin{eqnarray}
    \left. K_\mathrm{E}(r) \right|_{m = 0} \!\!\!\! &=& \!\!\!\! \left. K_\mathrm{S}(r) \right|_{m = 0} = \left. g_{rr} \sqrt{\frac{\left( L - q A_\varphi \right)^2}{g_{t\varphi}^2 - g_{tt} g_{\varphi\varphi}}} \right|_{\theta = \frac{\pi}{2}},
    \\
    \left. V_\mathrm{E}(r) \right|_{m = 0} \!\!\!\! &=& \!\!\!\! \left. V_\mathrm{S}(r) \right|_{m = 0} \; = \left. \frac{1}{g_{\varphi \varphi}} \sqrt{\left( L \! - \! q A_\varphi \right)^2 \left( g_{t\varphi}^2 \! - \! g_{tt} g_{\varphi\varphi} \right)} \! - \! L \frac{g_{t\varphi}}{g_{\varphi\varphi}} \! - \! q \left( A_t \! - \! \frac{g_{t\varphi}}{g_{\varphi\varphi}} A_\varphi \right) \right|_{\theta = \frac{\pi}{2}}.
\end{eqnarray}
    For a massless particle, the dilaton contribution vanishes, which eliminates the frame dependence. Consequently, the Lyapunov exponents in the Einstein and string frames coincide, as shown in
\begin{equation}
    \left. \lambda_\mathrm{E}^2 \right|_{m = 0} = \left. \lambda_\mathrm{S}^2 \right|_{m = 0}.
\end{equation}    
    In the subsequent analysis, the massless case is excluded because the Lyapunov exponent exhibits no frame dependence. The bound on chaos is considered in the form originally expressed as $\lambda \leq \kappa$, where $\kappa$ is the temperature of the system~\cite{Maldacena:2015waa} and surface gravity~\cite{Hashimoto:2016dfz}, respectively. For convenience, in the analysis, we employ the squared form $\lambda^2 \leq \kappa^2$ and introduce the quantity $\Delta^2$, referred to as the bound on chaos, which measures the difference between $\kappa^2$ and $\lambda^2$. The definition is given by
\begin{equation}
    \Delta^2 \equiv \kappa^2 - \lambda^2.
    \label{eq:boudn_on_chaos}
\end{equation}
    A non-negative value $(\Delta^2 \geq 0)$ indicates that the bound on chaos is satisfied, whereas a negative value $(\Delta^2 < 0)$ indicates that the bound is violated.

\section{Dilatonic Reissner--Nordstr\"om--AdS Black Hole} \label{sec_RN}
    In EMD theory, the four-dimensional effective action in the Einstein frame is given by
\begin{equation}
    S_\mathrm{E} = \int \mathrm{d}^4x \sqrt{-g} \left\{ R - \frac{1}{2} \partial_\mu \phi \partial^\mu \phi - e^{-\alpha \phi} F_{\mu\nu} F^{\mu\nu} - U(\phi, \Lambda) \right\},
\end{equation}
\noindent where $R$ is the Ricci scalar, $\phi$ is the dilaton field, $F_{\mu \nu}$ is the Maxwell field strength nonminimally coupled to the dilaton field with a coupling constant $\alpha$, and $U(\phi)$ is the potential of the dilaton including the cosmological constant $\Lambda$~\cite{Gao:2005xv}, which is defined by
\begin{equation}
    U(\phi) = \frac{2 \Lambda}{3 \left( 1 + \alpha^2 \right)^2} \left[ \alpha^2 \left( 3 \alpha^2 - 1 \right) e^{- \phi / \alpha} + \left( 3 - \alpha^2 \right) e^{\alpha \phi} + 8 \alpha^2 e^{\frac{1}{2} \left( \alpha \phi - \phi / \alpha \right)} \right].
\end{equation}
    The most general form of the static and spherically symmetric metric in the Einstein frame is given by
\begin{equation}
    \mathrm{d}s_\mathrm{E}^2 = - f(r) \mathrm{d}t^2 + \frac{\mathrm{d}r^2}{g(r)} + \mathcal{R}^2(r) \left( \mathrm{d}\theta^2 + \sin^2\theta \mathrm{d}\varphi^2 \right),
\end{equation}
\noindent and the solution for the dilatonic RN black hole is expressed as
\begin{eqnarray}
    f(r) &=& \left( 1 - \frac{r_+}{r} \right) \left( 1 - \frac{r_-}{r} \right)^\frac{1 - \alpha^2}{1 + \alpha^2} - \frac{\Lambda}{3} r^2 \left( 1 - \frac{r_-}{r} \right)^\frac{2 \alpha^2}{1 + \alpha^2},
    \\
    g(r) &=& f(r),
    \\
    \mathcal{R}(r) &=&  r \left( 1 - \frac{r_-}{r} \right)^\frac{\alpha^2}{1 + \alpha^2},
    \\
    \phi(r) &=& \frac{2 \alpha}{1 + \alpha^2} \log\left( 1 - \frac{r_-}{r} \right).
\end{eqnarray}
    The quantities $r_+$ and $r_-$ are positive values determined by the mass $M$ and electric charge $Q$ of the black hole. These quantities satisfy the following relationships:
\begin{equation}
    2 M = r_+ + \frac{1 - \alpha^2}{1 + \alpha^2} r_-, \qquad Q^2 = \frac{r_+ r_-}{1 + \alpha^2},
\end{equation}
\noindent which are equivalently expressed as
\begin{equation}
    r_+ = M + \sqrt{M^2 - \left( 1 - \alpha^2 \right) Q^2}, \quad r_- = \frac{\left( 1 + \alpha^2 \right) Q^2}{r_+}.
\end{equation}
    In the asymptotically flat limit $(\Lambda = 0)$, the quantities $r_+$ and $r_-$ correspond to the outer event horizon and the inner Cauchy horizon, respectively. Furthermore, when $\alpha = 1$, the solution is reduced to an electrically charged GMGHS black hole~\cite{Gibbons:1987ps, Garfinkle:1990qj}. The surface gravity is identical in both frames, and its general form is
\begin{equation}
    \kappa_\mathrm{E} = \kappa_\mathrm{S} = \frac{1}{2 r_\mathrm{h}^2}\left[ 3 r_+ - 2 r_\mathrm{h} - r_- - \frac{4\left( r_+ - r_\mathrm{h} \right) r_-}{\left( 1 + \alpha^2 \right) r_\mathrm{h}} \right] \left( 1 - \frac{r_-}{r_\mathrm{h}} \right)^{- \frac{2 \alpha^2}{1 + \alpha^2}}.
    \label{eq:surface_grvity}
\end{equation}  
\noindent where $r_\mathrm{h}$ denotes the outer event horizon, determined by $f(r_\mathrm{h}) = 0$. The kinetic term~\eqref{eq:string_kinetic_term} and effective potential~\eqref{eq:string_effective_potential} in both frames are given by
\begin{eqnarray}
    K_\mathrm{E}(r) \!\!\!\!\! &=& \!\!\!\!\! \frac{\left( 1 \! - \! \frac{r_-}{r} \right)^{\! \frac{\alpha^2 - 6}{2 \left( 1 + \alpha^2 \right)}} \!\! \sqrt{\frac{L^2}{r^2} \! + \! m^2 \! \left( 1 \! - \! \frac{r_-}{r} \right)^{\! \frac{2 \alpha^2}{1 + \alpha^2}}}}{\left[ 1 - \frac{r_+}{r} - \frac{\Lambda}{3} r^2 \left( 1 - \frac{r_-}{r} \right)^{- \frac{1 - 2 \alpha^2}{1 + \alpha^2}} \right]^{3/2}}, \, K_\mathrm{S}(r) \! = \! \frac{\left( 1 \! - \! \frac{r_-}{r} \right)^{\! \frac{\alpha^2 - 6}{2 \left( 1 + \alpha^2 \right)}} \!\! \sqrt{\frac{L^2}{r^2} \! + \! m^2 \! \left( 1 \! - \! \frac{r_-}{r} \right)^{\! \frac{2 \alpha \left( 1 +  \alpha \right)}{1 + \alpha^2}}}}{\left[ 1 - \frac{r_+}{r} - \frac{\Lambda}{3} r^2 \left( 1 - \frac{r_-}{r} \right)^{- \frac{1 - 2 \alpha^2}{1 + \alpha^2}} \right]^{3/2}},
    \\
    V_\mathrm{E}(r) \!\!\!\!\! &=& \!\!\!\!\! \frac{q Q}{r} + \left( 1 - \frac{r_-}{r} \right)^\frac{1 - 3 \alpha^2}{2 \left( 1 + \alpha^2 \right)} \sqrt{1 - \frac{r_+}{r} - \frac{\Lambda}{3} r^2 \left( 1 - \frac{r_-}{r} \right)^{- \frac{1 - 3 \alpha^2}{1 + \alpha^2}}} \sqrt{\frac{L^2}{r^2} + m^2 \left( 1 - \frac{r_-}{r} \right)^\frac{2 \alpha^2}{1 + \alpha^2}},
    \\
    V_\mathrm{S}(r) \!\!\!\!\! &=& \!\!\!\!\! \frac{q Q}{r} + \left( 1 - \frac{r_-}{r} \right)^\frac{1 - 3 \alpha^2}{2 \left( 1 + \alpha^2 \right)} \sqrt{1 - \frac{r_+}{r} - \frac{\Lambda}{3} r^2 \left( 1 - \frac{r_-}{r} \right)^{- \frac{1 - 2 \alpha^2}{1 + \alpha^2}}} \sqrt{\frac{L^2}{r^2} + m^2 \left( 1 - \frac{r_-}{r} \right)^\frac{2 \alpha \left( 1 + \alpha \right)}{1 + \alpha^2}}.
\end{eqnarray}
    The electric charge of the particle, $q$, that satisfies the condition $V'(r_0) = 0$ at the local maximum $r_0$ in both frames is determined by
\begin{equation}
    \left. q_\mathrm{E} \right|_{r = r_0} = \frac{A_\mathrm{E}}{B_\mathrm{E}}, \quad \left. q_\mathrm{S} \right|_{r = r_0} = \frac{A_\mathrm{S}}{B_\mathrm{S}},
    \label{eq:q}
\end{equation}
    where
\begin{eqnarray}
    A_\mathrm{E} \!\!\! &=& \!\!\! m^2 \left( 1 - \frac{r_-}{r_0} \right)^{- \frac{1}{2}} \left[ \left( 1 + \alpha^2 - \frac{2 r_-}{r_0} \right) r_+ - \left( 1 - \alpha^2 \right) r_- - \frac{2 \Lambda}{3} r_0^3 \left( 1 + \alpha^2 - \frac{r_-}{r_0} \right) \left( 1 - \frac{r_-}{r_0} \right)^{{- \frac{1 - 3 \alpha^2}{1 + \alpha^2}}} \right] \nonumber
    \\
    && + \frac{L^2}{r^2} \left( 1 - \frac{r_-}{r_0} \right)^{- \frac{1 + 5 \alpha^2}{2 \left( 1 + \alpha^2 \right)}} \left[ 3 \left( 1 + \alpha^2 - \frac{4 r_-}{3 r_0} \right) r_+ + \left( 3 - \alpha^2 \right) r_- - 2 \left( 1 + \alpha^2 \right) r_0 \right],
    \\
    B_\mathrm{E} \!\!\! &=& \!\!\! 2 \left( 1 + a^2 \right) Q \sqrt{ 1 - \frac{r_+}{r_0} - \frac{\Lambda}{3} r_0^2 \left( 1 - \frac{r_-}{r_0} \right)^{- \frac{1 - 3 \alpha^2}{1 + \alpha^2}}} \sqrt{\frac{L^2}{r_0^2} + m^2 \left( 1 - \frac{r_-}{r_0} \right)^\frac{2 \alpha^2}{1 + \alpha^2}},
    \\
    A_\mathrm{S} \!\!\! &=& \!\!\! m^2 \left( 1 - \frac{r_-}{r_0} \right)^{- \frac{1 - 4 \alpha + \alpha^2}{2 \left( 1 + \alpha^2 \right)}} \left\{ \left[ 1 + \alpha^2 + \frac{2 \left( 1 + \alpha \right) r_-}{r} \right] r_+ - \left( 1 - \alpha^2 \right) r_- \right. \nonumber
    \\
    && \quad \left. - \frac{2 \Lambda}{3} r_0^3 \left[ 1 + \alpha^2 - \left( 1 - \alpha \right) \frac{r_-}{r_0} \right] \left( 1 - \frac{r_-}{r_0} \right)^{- \frac{1 - 3 \alpha^2}{1 + \alpha^2}} \right\} \nonumber
    \\
    && + \frac{L^2}{r^2} \left( 1 - \frac{r_-}{r_0} \right)^{- \frac{1 + 6 \alpha^2}{2 \left( 1 + \alpha^2 \right)}} \left[ 3 \left( 1 + \alpha^2 - \frac{4 r_-}{3 r_0} \right) r_+ + \left( 3 - \alpha^2 \right) r_- - 2 \left( 1 + \alpha^2 \right) r_0 \right],
    \\
    B_\mathrm{S} \!\!\! &=& \!\!\! 2 \left( 1 + a^2 \right) Q \sqrt{ 1 - \frac{r_+}{r_0} - \frac{\Lambda}{3} r_0^2 \left( 1 - \frac{r_-}{r_0} \right)^{- \frac{1 - 2 \alpha^2}{1 + \alpha^2}}} \sqrt{\frac{L^2}{r_0^2} + m^2 \left( 1 - \frac{r_-}{r_0} \right)^\frac{2 \alpha \left( 1 + \alpha \right)}{1 + \alpha^2}}.
\end{eqnarray}

\subsection{Asymptotically Flat Spacetime}
    The analysis begins with an asymptotically flat dilatonic RN black hole. In the asymptotically flat limit, the outer horizon $r_\mathrm{h}$ coincides with $r_+$, and the inner horizon $r_-$ is given by
\begin{equation}
    r_\mathrm{h} = r_+ = M + \sqrt{M^2 - \left( 1 - \alpha^2 \right) Q^2}, \quad r_- = \frac{\left( 1 + \alpha^2 \right) Q^2}{r_+}.
\end{equation}

\subsubsection{Extremal Case}
    The extremal case corresponds to the configuration in which the outer and inner horizons coincide $( r_+ = r_- )$. In the extremal configuration, the surface gravity~\eqref{eq:surface_grvity} in both frames is given by
\begin{equation}
    \left. \kappa_\mathrm{E} \right|_{\Lambda = 0} = \left. \frac{\left( 1 - \frac{r_\mathrm{h}}{r} \right)^\frac{1 - \alpha^2}{1 + \alpha^2}}{\left( 1 + \alpha^2 \right) r_\mathrm{h}} \right|_{r = r_\mathrm{h}}, \qquad \left. \kappa_\mathrm{S} \right|_{\Lambda = 0} = \left. \frac{\left( 1 + \alpha \right) \left( 1 - \frac{r_\mathrm{h}}{r} \right)^\frac{1 - \alpha^2}{1 + \alpha^2}}{\left( 1 + \alpha^2 \right) r_\mathrm{h}} \right|_{r = r_\mathrm{h}},
\end{equation}
    The surface gravities $\kappa_\mathrm{E}$ and $\kappa_\mathrm{S}$ depend on the dilaton coupling constant $\alpha$ and are classified according to the range of $\alpha$ as shown in
\begin{equation}
\begin{aligned}
    \left. \kappa_\mathrm{E} \right|_{\Lambda = 0} =
    \begin{cases}
        0 & \alpha^2 < 1
        \\
        \frac{1}{2 r_\mathrm{h}} & \alpha^2 = 1
    \end{cases},
    \qquad
    & \left. \kappa_\mathrm{S} \right|_{\Lambda = 0} =
    \begin{cases}
        0 & \alpha^2 < 1
        \\
        \frac{1 + \alpha}{2 r_\mathrm{h}} & \alpha^2 = 1
    \end{cases}.
\end{aligned}
\end{equation}
    For $\alpha^2 < 1$, the surface gravity vanishes in both frames, corresponding to an extremal black hole with zero temperature. When $\alpha^2 = 1$, the two frames exhibit a distinct difference in surface gravity, which makes it appropriate to present the corresponding line elements explicitly. In the Einstein frame, the line element and surface gravity are reduced to
\begin{equation}
    \left. \mathrm{d}s_\mathrm{E}^2 \right|_{\alpha^2 = 1} = - \left( 1 - \frac{r_\mathrm{h}}{r} \right) \mathrm{d}t^2 + \left( 1 - \frac{r_\mathrm{h}}{r} \right)^{-1} \mathrm{d}r^2 + r \left( r - r_\mathrm{h} \right) \left( \mathrm{d}\theta^2 + \sin^2\theta \mathrm{d}\varphi^2 \right), \quad \left. \kappa_\mathrm{E} \right|_{\alpha^2 = 1} = \frac{1}{2 r_\mathrm{h}}.
\end{equation}
    Because the black hole possesses a single event horizon, it cannot be regarded as an extremal black hole, and its surface gravity remains nonzero in the Einstein frame. By contrast, in the string frame for $\alpha = -1$, the line element and surface gravity take the form
\begin{equation}
    \left. \mathrm{d}s_\mathrm{S}^2 \right|_{\alpha = -1} = - \mathrm{d}t^2 + \left( 1 - \frac{r_\mathrm{h}}{r} \right)^{-2} \mathrm{d}r^2 + r^2 \left( \mathrm{d}\theta^2 + \sin^2\theta \mathrm{d}\varphi^2 \right), \quad \left. \kappa_\mathrm{S} \right|_{\alpha = -1} = 0.
\end{equation}
    The metric exhibits a degenerate horizon at $r = r_\mathrm{h}$, indicating that the black hole is extremal and that the surface gravity vanishes. Because $g_{tt} = -1$, the degenerate horizon fails to be a Killing horizon, and the standard definition of surface gravity is not applicable. Applying Kodama--Hayward surface gravity~\cite{Kodama:1979vn, Hayward:1997jp} yields $\kappa = 0$, as expected for an extremal black hole. For $\alpha = 1$ in the string frame, the line element and surface gravity are given by
\begin{equation}
    \left. \mathrm{d}s_\mathrm{S}^2 \right|_{\alpha = 1} = - \left( 1 - \frac{r_\mathrm{h}}{r} \right)^2 \mathrm{d}t^2 + \mathrm{d}r^2 + \left( r - r_\mathrm{h} \right)^2 \left( \mathrm{d}\theta^2 + \sin^2\theta \mathrm{d}\varphi^2 \right), \quad \left. \kappa_\mathrm{S} \right|_{\alpha = 1} = \frac{1}{r_\mathrm{h}}.
\end{equation}
    No event horizon exists in this configuration, indicating that the solution corresponds to a naked singularity. The Killing horizon and surface gravity are formally defined at $r = r_\mathrm{h}$, although the absence of an event horizon precludes a direct association with the Hawking temperature, and the analysis for this case is consequently not performed. As shown, the surface gravity differs between frames in the limiting case of $\alpha^2 = 1$, and the bound on chaos is expected to adjust accordingly. In the following analysis, we examine the modification of the bound on chaos under various limiting cases.
    
    \noindent \textbf{\textit{Vanishing angular momentum}}---For vanishing angular momentum, the bound on chaos~\eqref{eq:boudn_on_chaos} in both frames reduces to
\begin{eqnarray}
    \left. \Delta_\mathrm{E}^2 \right|_{\{ \Lambda = 0, \, L = 0 \}} &=&
    \begin{cases}
        - \frac{\alpha^2 r_\mathrm{h}^2}{\left( 1 + \alpha^2 \right)^2 r_0^4} \left( 1 - \frac{r_\mathrm{h}}{r_0} \right)^\frac{2 ( 1 - \alpha^2 )}{1 + \alpha^2} & \left( \alpha^2 < 1 \right)
        \\
        \frac{1}{4 r_\mathrm{h}^2} \left( 1 - \frac{r_\mathrm{h}^4}{r_0^4} \right) & \left( \alpha^2 = 1 \right)
    \end{cases},
    \label{eq:boudn_on_chaos_extreme_L=0_Einstein}
    \\
    \nonumber
    \\
    \left. \Delta_\mathrm{S}^2 \right|_{\{ \Lambda = 0, \, L = 0 \}} &=&
    \begin{cases}
        - \frac{\alpha \left( \alpha^2 - 1 \right) r_\mathrm{h}^2}{\left( 1 + \alpha^2 \right)^2 r_0^4} \left( 1 - \frac{r_\mathrm{h}}{r_0} \right)^\frac{2 ( 1 - \alpha^2 )}{1 + \alpha^2} & \left( \alpha^2 < 1 \right)
        \\
        0 & \left( \alpha = -1 \right)
    \end{cases}.
    \label{eq:boudn_on_chaos_extreme_L=0_String}
\end{eqnarray}
    Depending on the value of $\alpha$, the corresponding signs of the bounds on chaos~\eqref{eq:boudn_on_chaos_extreme_L=0_Einstein}~and~\eqref{eq:boudn_on_chaos_extreme_L=0_String} are obtained as follows:
\begin{equation}
\begin{aligned}
    \left. \Delta_\mathrm{E}^2 \right|_{\{ \Lambda = 0, \, L = 0 \}}
    \begin{cases}
        \geq 0 & \left( \alpha = -1, \; 0, \; 1 \right)
        \\
        < 0 & \left( 0 < \alpha^2 < 1 \right)
    \end{cases},
    \qquad
    & \left. \Delta_\mathrm{S}^2 \right|_{\{ \Lambda = 0, \, L = 0 \}}
    \begin{cases}
        \geq 0 & \left( \alpha = -1, \; 0 \leq \alpha < 1 \right)
        \\
        < 0 & \left( -1 < \alpha < 0 \right) 
    \end{cases}.
\end{aligned}
\end{equation}
    A violation of the bound on chaos occurs in the Einstein frame for $0 < \alpha^2 < 1$, whereas in the string frame, it is constrained to $-1 < \alpha < 0$. Moreover, marginal cases with $\Delta^2 = 0$ occur in the string frame at $\alpha = -1$ and $\alpha = 0$, while in the Einstein frame, the marginal case appears only at $\alpha = 0$.

    \noindent \textbf{\textit{Large angular momentum}}---In the large angular momentum limit, the electric charge of the particle~\eqref{eq:q} takes an identical form in both frames and reduces to
\begin{equation}
    \frac{q}{L} = \mp \frac{\left( 1 + \alpha^2 \right) r_0 - 2 r_\mathrm{h}}{r_0 r_\mathrm{h} \sqrt{1 + \alpha^2}} \left( 1 - \frac{r_\mathrm{h}}{r_0} \right)^{- \frac{2 \alpha^2}{1 + \alpha^2}} + \mathcal{O}(L^{-2}),
\end{equation}
\noindent where the upper sign corresponds to a large positive angular momentum $(L \gg 0)$ and the lower sign to a large negative angular momentum $(L \ll 0)$. The local maximum point $r_0$ is given by
\begin{equation}
    r_0 = \frac{2 r_\mathrm{h}}{1 + \alpha^2}.
\end{equation}
    Because $r_0 > r_\mathrm{h}$, a necessary condition for the occurrence of the local maximum point $r_0$ is $\alpha^2 < 1$. The bound on chaos~\eqref{eq:boudn_on_chaos} is identical in the Einstein and string frames and takes the form
\begin{equation}
    \left. \Delta_\mathrm{E}^2 \right|_{\{ \Lambda = 0, \, L \to \pm \infty \}} = \left. \Delta_\mathrm{S}^2 \right|_{\{ \Lambda = 0, \, L \to \pm \infty \}} = - \frac{\left( 1 + \alpha^2 \right)^2}{2 \left( 1 - \alpha^2 \right) r_\mathrm{h}^2} \left( \frac{1 - \alpha^2}{2} \right)^\frac{4}{1 + \alpha^2} + \mathcal{O}(L^{-2}) < 0,
\end{equation}
\noindent indicating a violation of the bound on chaos in both frames.
    
    \noindent \textbf{\textit{Near-horizon limit}}---The analysis considers the near-horizon limit, defined by $r_0 = r_\mathrm{h} + \epsilon$, where $\epsilon$ represents a small deviation. In the near-horizon regime, the bound on chaos~\eqref{eq:boudn_on_chaos} in both frames is reduced to
\begin{eqnarray}
    \left. \Delta_\mathrm{E}^2 \right|_{\{ \Lambda = 0, \, r_0 = r_\mathrm{h} + \epsilon \}} &=&
    \begin{cases}
        - \frac{2 L^2}{\left( L^2 + m^2 r_\mathrm{h}^2 \right) r_\mathrm{h}^5} \epsilon^3 + \mathcal{O}(\epsilon^4) & \quad \;\;\; \left( \alpha = 0 \right)
        \\
        - \frac{2 \alpha^2 \left( 1 - \alpha^2 \right)}{\left( 1 + \alpha^2 \right)^2} r_\mathrm{h}^{- \frac{4}{1 + \alpha^2}} \epsilon^\frac{2 \left( 1 - \alpha^2 \right)}{1 + \alpha^2} + \mathcal{O}(\epsilon^\frac{2}{1 + \alpha^2}) & \quad \;\;\; \left( 0 < \alpha^2 < 1 \right)
        \\
        \frac{1}{4 r_\mathrm{h}^2} + \mathcal{O}(\epsilon^2) & \quad \;\;\; \left( \alpha^2 = 1 \right)
    \end{cases},
    \label{eq:boudn_on_chaos_extreme_near_horizon_Einstein}
    \\
    \nonumber
    \\
    \left. \Delta_\mathrm{S}^2 \right|_{\{ \Lambda = 0, \, r_0 = r_\mathrm{h} + \epsilon \}} &=&
    \begin{cases}
        \frac{L^2 m^2}{\left( L^2 + m^2 r_\mathrm{h}^2 \right)^2 r_\mathrm{h}^2} \epsilon^2 + \mathcal{O}(\epsilon^3) & \left( \alpha = -1 \right)
        \\
        - \frac{\alpha \left( \alpha^2 - 1 \right)}{\left( 1 + \alpha^2 \right)^2} r_\mathrm{h}^{- \frac{4}{1 + \alpha^2}} \epsilon^\frac{2 \left( 1 - \alpha^2 \right)}{1 + \alpha^2} + \mathcal{O}(\epsilon^\frac{2 \left( 1 + \alpha \right) \left( 1 - 2 \alpha \right)}{1 + \alpha^2}) & \left( -1 < \alpha < 0 \right)
        \\
        - \frac{2 L^2}{\left( L^2 + m^2 r_\mathrm{h}^2 \right) r_\mathrm{h}^5} \epsilon^3 + \mathcal{O}(\epsilon^4) & \left( \alpha = 0 \right)
        \\
        - \frac{2 \alpha^2 \left( 1 - \alpha^2 \right)}{\left( 1 + \alpha^2 \right)^2} r_\mathrm{h}^{- \frac{4}{1 + \alpha^2}} \epsilon^\frac{2 \left( 1 - \alpha^2 \right)}{1 + \alpha^2} + \mathcal{O}(\epsilon^\frac{3 - \alpha^2}{1 + \alpha^2}) & \left( 0 < \alpha < 1 \right)
    \end{cases}.
    \label{eq:boudn_on_chaos_extreme_near_horizon_String}
\end{eqnarray}
    The signs of the bound on chaos~\eqref{eq:boudn_on_chaos_extreme_near_horizon_Einstein}~and~\eqref{eq:boudn_on_chaos_extreme_near_horizon_String} in the near-horizon limit are given by
\begin{equation}
\begin{aligned}
    \left. \Delta_\mathrm{E} \right|_{\{ \Lambda = 0, \, r_0 = r_\mathrm{h} + \epsilon \}}
    \begin{cases}
        > 0 & \left( \alpha^2 = 1 \right)
        \\
        < 0 & \left( \alpha^2 < 1 \right)
    \end{cases},
    \qquad
    & \left. \Delta_\mathrm{S} \right|_{\{ \Lambda = 0, \, r_0 = r_\mathrm{h} + \epsilon \}}
    \begin{cases}
        > 0 & \left( \alpha = -1 \right)
        \\
        < 0 & \left( \alpha^2 < 1 \right)
    \end{cases}.
\end{aligned}
\end{equation}
    Although the values of the bound on chaos differ between the Einstein and string frames, violations occur in both frames for $\alpha^2 < 1$.

\subsubsection{Non-Extremal Case}
    This analysis considers the non-extremal case, which is characterized by distinct outer and inner horizons $( r_\mathrm{h} > r_- )$. In the non-extremal regime, the surface gravities~\eqref{eq:surface_grvity} in both frames are reduced to
\begin{equation}
    \left. \kappa_\mathrm{E} \right|_{\Lambda = 0} = \left. \kappa_\mathrm{S} \right|_{\Lambda = 0} = \frac{1}{2 \left( r_\mathrm{h} - r_- \right)} \left( 1 - \frac{r_-}{r_\mathrm{h}} \right)^\frac{2}{1 + \alpha^2}.
\end{equation}
    
    \noindent \textbf{\textit{Vanishing angular momentum in near-horizon limit}}---In the near-horizon limit with vanishing angular momentum, the bound on chaos~\eqref{eq:boudn_on_chaos} in both frames takes the form
\begin{eqnarray}
    \left. \Delta_\mathrm{E}^2 \right|_{\{ \Lambda = 0, \, L = 0, \, r_0 = r_\mathrm{h} + \epsilon \}} &=& \frac{1}{r_\mathrm{h}^3} \left( 1 - \frac{r_-}{r_\mathrm{h}} \right)^\frac{2 \left( 1 - \alpha^2 \right)}{1 + \alpha^2} \epsilon + \mathcal{O}(\epsilon^2) > 0,
    \\
    \left. \Delta_\mathrm{S}^2 \right|_{\{ \Lambda = 0, \, L = 0, \, r_0 = r_\mathrm{h} + \epsilon \}} &=& \frac{1}{r_\mathrm{h}^3} \left[ 1 - \frac{ 1 - \alpha + \alpha^2}{1 + \alpha^2} \frac{r_-}{r_\mathrm{h}} \right] \left( 1 - \frac{r_-}{r_\mathrm{h}} \right)^\frac{1 - 3 \alpha^2}{1 + \alpha^2} \epsilon + \mathcal{O}(\epsilon^2)
\end{eqnarray}
    The positive value of $\Delta_\mathrm{E}^2$ indicates that in the Einstein frame, the bound on chaos is satisfied. On the other hand, in the string frame, the bound is negative when
\begin{equation}
     r_\mathrm{h} < \left( 1 - \frac{\alpha}{1 + \alpha^2} \right) r_-, \qquad \alpha < 0,
\end{equation}
    resulting in
\begin{equation}
    \left. \Delta_\mathrm{S}^2 \right|_{\{ \Lambda = 0, \, L = 0, \, r_0 = r_\mathrm{h} + \epsilon \}} < 0,
\end{equation}
    which corresponds to a violation of the bound on chaos for the specified parameter range in the string frame.

\subsection{Asymptotically AdS Spacetime}
    The analysis then proceeds to the dilatonic RN--AdS black hole. In the presence of a negative cosmological constant $(\Lambda < 0)$, the outer horizon is denoted by $r_\mathrm{h}$ and satisfies
\begin{equation}
    r_- < r_\mathrm{h} < r_+,
\end{equation}
\noindent where $r_-$ corresponds to the inner horizon, whereas $r_+$ no longer represents the outer horizon as in the asymptotically flat case.

\subsubsection{Extremal Case}
\begin{figure}[H]
    \centering
    \includegraphics[width=14.0cm]{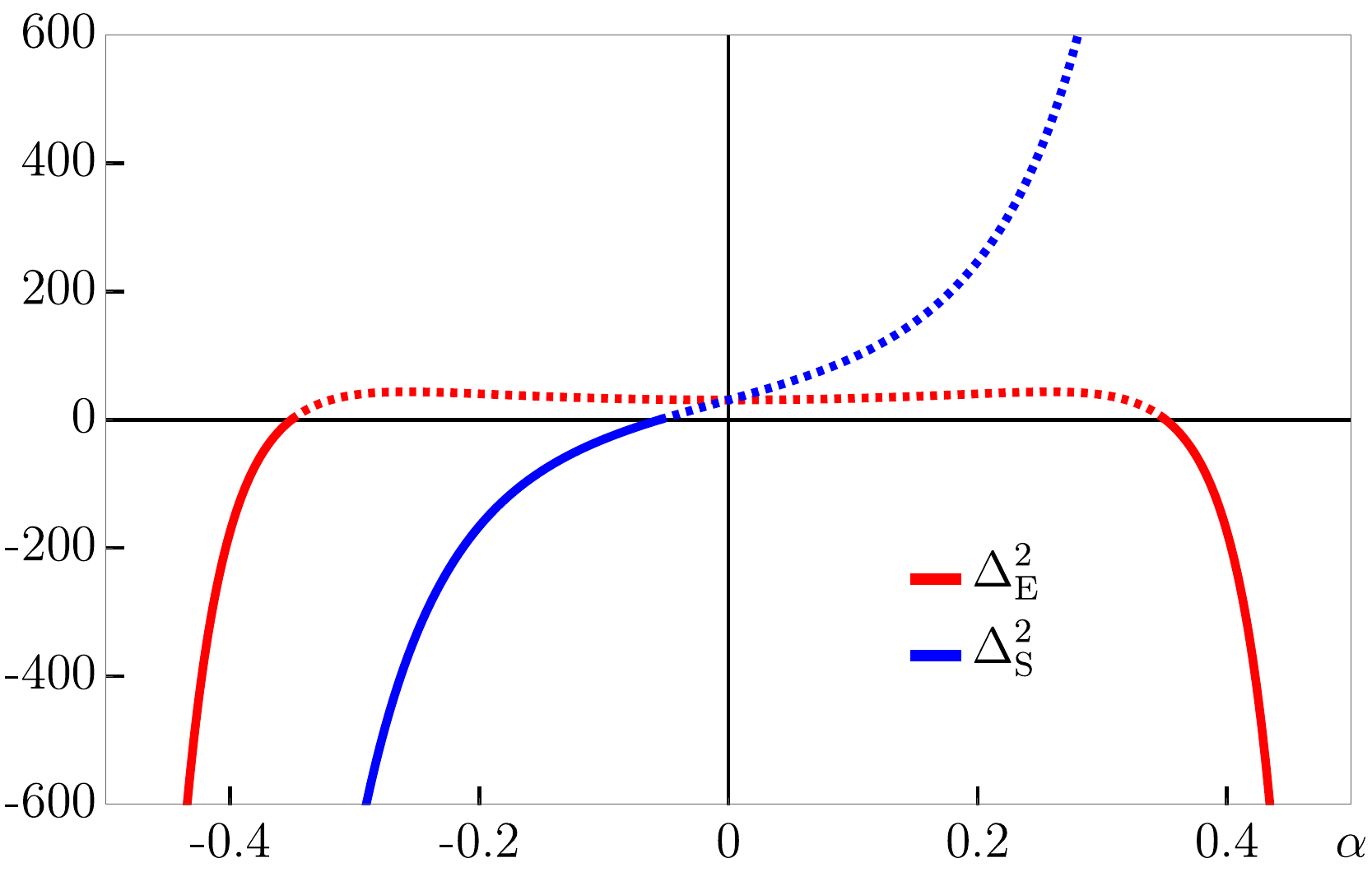}
    \caption{Bound on chaos in the near-horizon limit with vanishing angular momentum of extremal dilatonic Reissner--Nordstr\"om--AdS black hole backgrounds.}
    \label{fig:DRN_AdS_Ext_NH}
\end{figure}

    The extremal configuration for the dilatonic RN--AdS black hole is characterized by the relations
\begin{equation}
    r_+ = \frac{r_\mathrm{h} \left[ \left( 3 - \alpha^2 \right) r_- - 2 \left( 1 + \alpha^2 \right) r_\mathrm{h} \right]}{4 r_- - 3 \left( 1 + \alpha^2 \right) r_\mathrm{h}}, \quad \Lambda = - \frac{3 \left( 1 + \alpha^2 \right)}{\left[ 3 \left( 1 + \alpha^2 \right) r_\mathrm{h} - 4 r_- \right] r_\mathrm{h}} \left( 1 - \frac{r_-}{r_\mathrm{h}} \right)^\frac{2 \left( 1 - \alpha^2 \right)}{1 + \alpha^2}.
\end{equation}
    An AdS spacetime with $\Lambda < 0$ is ensured when the following conditions are satisfied:
\begin{equation}
    r_- < r_\mathrm{h} < \frac{4}{3 \left( 1 + \alpha^2 \right)} r_-, \quad \alpha^2 < \frac{1}{3}.
\end{equation}
    The surface gravity vanishes in both frames, indicating an extremal black hole, which is expressed as
\begin{equation}
    \kappa_\mathrm{E} = \kappa_\mathrm{S} = 0.
\end{equation}
    \noindent \textbf{\textit{Vanishing angular momentum in near-horizon limit}}---For vanishing angular momentum in the near-horizon limit, the bound on chaos~\eqref{eq:boudn_on_chaos} in both frames reduces to
\begin{eqnarray}
    \left. \Delta_\mathrm{E} \right|_{\{ L = 0, \, r_0 = r_\mathrm{h} + \epsilon \}} \!\!\! &=& \!\!\! \left[ 2 \left( \frac{3}{\alpha^2} + 5 - 2 \alpha^2 \right) \delta^3 + 6 \left( 1 + 5 \alpha^2 \right) \delta^2 + 3 \left( 1 - 4 \alpha^2 + 3 \alpha^4 \right) \delta - 1 + \left( 4 - 3 \alpha^2 \right) \alpha^2 \right] \nonumber
    \\
    && \times \frac{2 \alpha^2 \left[ 2 \left( 3 - \alpha^2 \right) \delta^2 - 4 \left( 1 - 3 \alpha^2 \right) \delta + 1 - \left( 4 - 3 \alpha^2 \right) \alpha^2 \right]}{3 \left( 1 + \alpha^2 \right)^3 \left( 4 \delta + 3 \alpha^2 - 1 \right)^2 r_h^5} \delta^{- \frac{1 + 5 \alpha^2}{1 + \alpha^2}} \epsilon^3 + \mathcal{O}(\epsilon^4),
    \\
    \left. \Delta_\mathrm{S} \right|_{\{ L = 0, \, r_0 = r_\mathrm{h} + \epsilon \}} \!\!\! &=& \!\!\! \left[ 2 \left( \frac{3}{\alpha^2} - \frac{9}{\alpha} + 5 + 3 \alpha - 2 \alpha^2 \right) \delta^3 + 6 \left( \frac{5}{\alpha} + 1 - 7 \alpha + 5 \alpha^2 \right) \delta^2 \right. \nonumber
    \label{eq:DRN_AdS_extreme_near_horizon_Einstein}
    \\
    && \left. - 3 \left( \frac{5}{\alpha} - 1 - 16 \alpha + 4 \alpha^2 + 3 \alpha^3 - 3 \alpha^4 \right) \delta + \frac{3}{\alpha} - 1 - 12 \alpha + 4 \alpha^2 + 9 \alpha^3 - 3 \alpha^4 \right] \nonumber
    \\
    && \times \frac{2 \alpha^2 \left[ 2 \left( 3 - \alpha^2 \right) \delta^2 - 4 \left( 1 - 3 \alpha^2 \right) \delta + 1 - \left( 4 - 3 \alpha^2 \right) \alpha^2 \right]}{3 \left( 1 + \alpha^2 \right)^3 \left( 4 \delta + 3 \alpha^2 - 1 \right)^2 r_h^5} \delta^{- \frac{1 + 5 \alpha^2}{1 + \alpha^2}} \epsilon^3 + \mathcal{O}(\epsilon^4),
    \label{eq:DRN_AdS_extreme_near_horizon_String}
\end{eqnarray}
\noindent where $\delta = 1 - \frac{r_-}{r_\mathrm{h}}$. Expressions~\eqref{eq:DRN_AdS_extreme_near_horizon_Einstein}~and~\eqref{eq:DRN_AdS_extreme_near_horizon_String} exhibit conditional violation, as shown numerically in Fig.~\ref{fig:DRN_AdS_Ext_NH}.
    Fig.~\ref{fig:DRN_AdS_Ext_NH} presents the bound on chaos in the near-horizon region of extremal dilatonic RN--AdS black holes with vanishing angular momentum in the Einstein and string frames, corresponding to~\eqref{eq:DRN_AdS_extreme_near_horizon_Einstein}~and~\eqref{eq:DRN_AdS_extreme_near_horizon_String}. The parameters are set to $M = 1$, $r_- = 0.5$, and $\epsilon = 1$. The parameter $\epsilon$, originally introduced as a small expansion parameter, is fixed at $1$ for simplicity. The horizontal axis represents the coupling constant $\alpha$, whereas the vertical axis represents the bound on chaos, $\Delta^2$. The red line indicates the bound in the Einstein frame, whereas the blue line indicates the bound in the string frame. Solid lines signify negative values of the bound on chaos, whereas dashed lines signify positive values. The bound in the Einstein frame is negative for $\alpha > 0.3508$ and $\alpha < -0.3508$. By comparison, the bound in the string frame is negative for all $\alpha < -0.0553$. For the extremal black hole, the surface gravity vanishes, and the Lyapunov exponent evaluated at the local maximum of the potential remains positive. The positive region of the bound on chaos is inconsistent with the Lyapunov exponent evaluated at the local maximum and reflects merely an artifact of the analytical expressions in~\eqref{eq:DRN_AdS_extreme_near_horizon_Einstein}~and~\eqref{eq:DRN_AdS_extreme_near_horizon_String}. Consequently, positive values are excluded in the numerical evaluation presented in Fig.~\ref{fig:DRN_AdS_Ext_NH}. In the Einstein frame, violations of the bound occur when $\alpha^2$ exceeds a critical value, whereas in the string frame, violations are limited to a single direction in the $\alpha$ parameter space.
    
\begin{figure}[H]
    \centering
    \includegraphics[width=14.0cm]{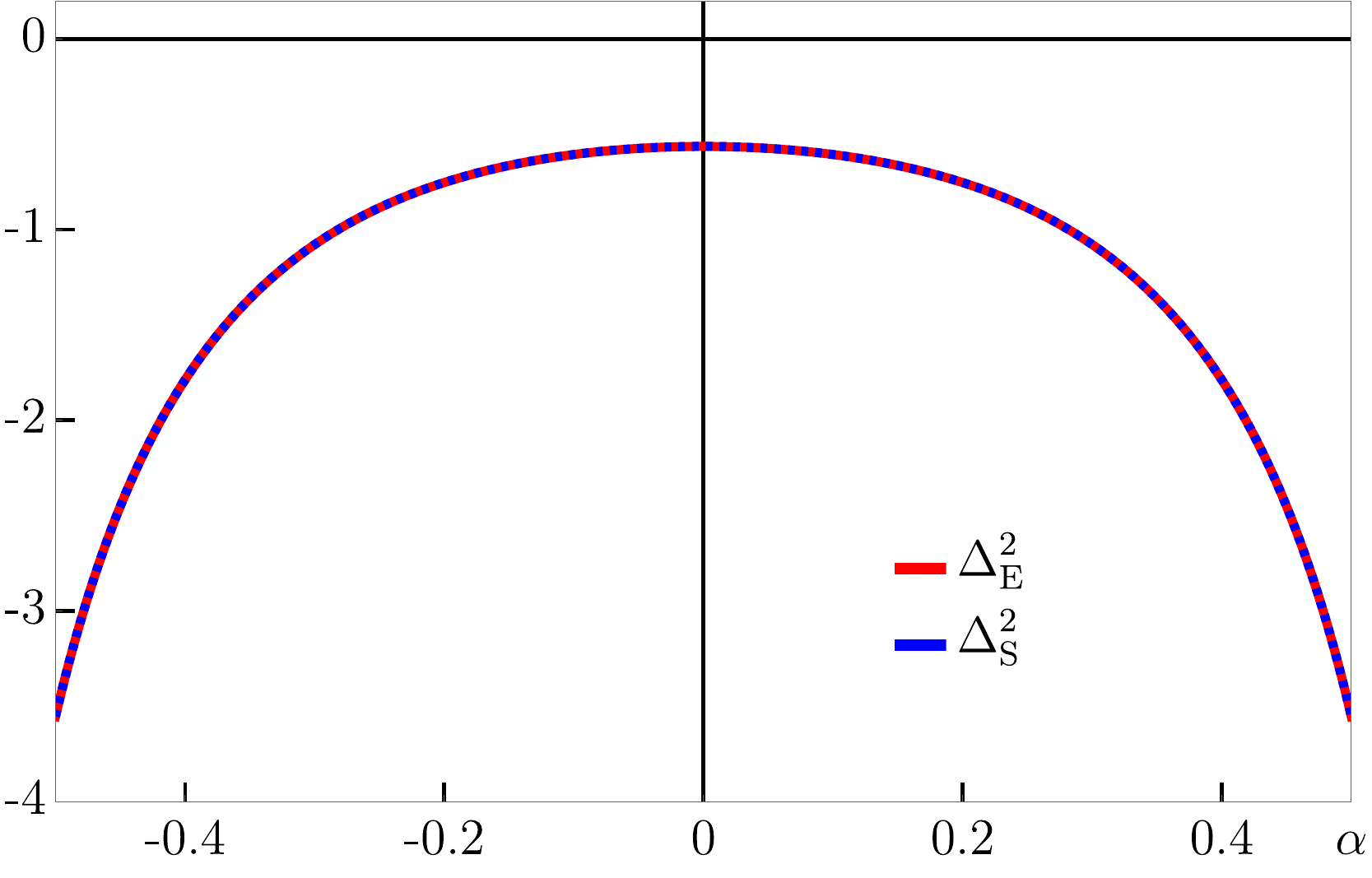}
    \caption{Bound on chaos in the large angular momentum limit of extremal dilatonic Reissner--Nordstr\"om--AdS black hole backgrounds.}
    \label{fig:DRN_AdS_Ext_LAM}
\end{figure}

    \noindent \textbf{\textit{Large angular momentum}}---In the large angular momentum limit, the electric charge of the particle~\eqref{eq:q} in both frames is given by
\begin{equation}
    \frac{q}{L} = \mp \frac{4 r_+ r_- - 3 \left( r_+ + r_- \right) r_0 + 2 r_0^2 + \alpha^2 \left( 2 r_0 + r_- - 3 r_+ \right) r_0}{2 Q \left( 1 + \alpha^2 \right) \ \sqrt{\left( 1 - \frac{r_+}{r_0} \right) \left( 1 - \frac{r_-}{r_0} \right)^\frac{1}{1 + \alpha^2} - \frac{\Lambda}{3} r^2 \left( 1 - \frac{r_-}{r_0} \right)^\frac{3 \alpha^2}{1 + \alpha^2}}} \left( 1 - \frac{r_-}{r_0} \right)^{- \frac{5 \alpha^2}{2 \left( 1 + \alpha^2 \right)}} + \mathcal{O}(L^{-2}),
\end{equation}
\noindent where the upper sign corresponds to a large positive angular momentum $(L \gg 0)$ and the lower sign to large negative angular momentum $(L \ll 0)$. The local maximum point $r_0$ is defined as follows:
\begin{equation}
    r_0 = 2 \left[ \frac{r_\mathrm{h} - r_-}{4 r_- - 3 \left( 1 + \alpha^2 \right) r_\mathrm{h}} + \frac{1}{1 + \alpha^2} \right] r_-.
\end{equation}
    The condition $r_0 > r_+$ restricts the local maximum point $r_0$ to cases where $\alpha^2 < 1$. The bound on chaos~\eqref{eq:boudn_on_chaos} is identical in the Einstein and string frames and is reduced to
\begin{equation}
    \left. \Delta_\mathrm{E}^2 \right|_{L \to \pm \infty} = \left. \Delta_\mathrm{S}^2 \right|_{L \to \pm \infty} = \mathcal{O}(1).
    \label{eq:DRN_AdS_extreme_large_angular_momentum}
\end{equation}
    Owing to the complicated form of the analytical expression~$\text{\eqref{eq:DRN_AdS_extreme_large_angular_momentum}}$, we illustrate its behavior numerically in Fig.~$\text{\ref{fig:DRN_AdS_Ext_LAM}}$.
    Fig.~\ref{fig:DRN_AdS_Ext_LAM} illustrates the bound on chaos in the large angular momentum limit of extremal dilatonic RN--AdS black holes in the Einstein and string frames, corresponding to~\eqref{eq:DRN_AdS_extreme_large_angular_momentum}. The parameters are set to $M = 1$ and $r_- = 0.5$. The horizontal axis represents $\alpha$, whereas the vertical axis represents the bound on chaos, $\Delta^2$. The red line indicates the bound in the Einstein frame, whereas the blue line indicates the bound in the string frame. In both frames, the bounds on chaos are identical, producing coinciding lines. Negative values in both frames denote violations in the large angular momentum limit.

\subsubsection{Non-Extremal Case}
    We extend the analysis to the non-extremal regime of the dilatonic RN--AdS black hole. In the non-extremal regime, the surface gravity in both frames is given by
\begin{equation}
    \kappa_\mathrm{E} = \kappa_\mathrm{S} = \frac{2 r_\mathrm{h} + \frac{4 r_+ r_-}{r_\mathrm{h}} - 3 \left( r_+ + r_- \right) + \alpha^2 \left( 2 r_\mathrm{h} + r_- - 3 r_+ \right)}{2 \left( 1 + \alpha^2 \right) r_\mathrm{h}^2} \left( 1 - \frac{r_-}{r_\mathrm{h}} \right)^{- \frac{2 \alpha^2}{1 + \alpha^2}}.
\end{equation}
    The cosmological constant is expressed in terms of the outer horizon $r_\mathrm{h}$ as follows:
\begin{equation}
    \Lambda = - \frac{3 \left( r_+ - r_\mathrm{h} \right)}{r_\mathrm{h}^3} \left( 1 - \frac{r_-}{r_\mathrm{h}} \right)^\frac{1 - 3 \alpha^2}{1 + \alpha^2}.
\end{equation}
    For the spacetime to remain AdS with $\Lambda < 0$, the condition is
\begin{equation}
    r_\mathrm{h} < r_+.
\end{equation}
    
    \noindent \textbf{\textit{Vanishing angular momentum in near-horizon limit}}---For a non-extremal configuration with vanishing angular momentum in the near-horizon region, the bound on chaos~\eqref{eq:boudn_on_chaos} is identical in both frames and is given by
\begin{eqnarray}
    \left. \Delta_\mathrm{E} \right|_{\{ L = 0, \, r_0 = r_\mathrm{h} + \epsilon \}} &=& \frac{\left[ 2 r_\mathrm{h} + \frac{4 r_+ r_-}{r_\mathrm{h}} - 3 \left( r_+ + r_- \right) + \alpha^2 \left( 2 r_\mathrm{h} + r_- - 3 r_+ \right) \right]^2}{\left( 1 + \alpha^2 \right)^2 r_\mathrm{h}^5} \left( 1 - \frac{r_-}{r_\mathrm{h}} \right)^{- \frac{4 \alpha^2}{1 + \alpha^2}} \epsilon \nonumber
    \\
    && \quad + \mathcal{O}(\epsilon^2) > 0,
    \\
    \left. \Delta_\mathrm{S} \right|_{\{ L = 0, \, r_0 = r_\mathrm{h} + \epsilon \}} &=& \frac{\left[ 2 r_\mathrm{h} + \frac{4 r_+ r_-}{r_\mathrm{h}} - 3 \left( r_+ + r_- \right) + \alpha^2 \left( 2 r_\mathrm{h} + r_- - 3 r_+ \right) \right]^2}{\left( 1 + \alpha^2 \right)^2 r_\mathrm{h}^5} \left( 1 - \frac{r_-}{r_\mathrm{h}} \right)^{- \frac{1 + 5 \alpha^2}{1 + \alpha^2}} \epsilon \nonumber
    \\
    && \quad \times \left[ 1 - \left( 1 - \frac{\alpha}{1 + \alpha^2} \right) \frac{r_-}{r_\mathrm{h}} \right] + \mathcal{O}(\epsilon^2).
\end{eqnarray}
    In the Einstein frame, the bound on chaos $\Delta_\mathrm{E}^2$ is satisfied, whereas in the string frame, violations occur when the parameters satisfy
\begin{equation}
    r_- < r_\mathrm{h} < \left( 1 - \frac{\alpha}{1 + \alpha^2} \right) r_-, \quad \alpha < 0,
\end{equation}
    leading to
\begin{equation}
    \left. \Delta_\mathrm{S} \right|_{\{ r_0 = r_+ + \epsilon, \, L = 0 \}} < 0,
\end{equation}
\noindent indicating that violations appear within a specific parameter range in the string frame.

    \noindent \textbf{\textit{Vanishing angular momentum in large negative cosmological constant limit}}---In the limit of a large negative cosmological constant with vanishing angular momentum, the electric charge of the particle~\eqref{eq:q} in both frames reduces to
\begin{eqnarray}
    \frac{q_\mathrm{E}}{\Lambda} &=& - \frac{m^2 r_0^2 \left[ \left( 1 + \alpha^2 \right) r_0 - r_- \right] \left( 1 - \frac{r_-}{r_0} \right)^{- \frac{1 - \alpha^2}{1 + \alpha^2}}}{\sqrt{3} \left( 1 + \alpha^2 \right) Q \sqrt{L^2 + m^2 r_0^2 \left( 1 - \frac{r_-}{r_0} \right)^\frac{2 \alpha^2}{1 + \alpha^2}}} \frac{1}{\sqrt{- \Lambda}},
    \label{eq:DRN_AdS_large_cosmological_constant_Einstein}
    \\
    \frac{q_\mathrm{S}}{\Lambda} &=& - \frac{m^2 r_0^2 \left[ \left( 1 + \alpha^2 \right) r_0 - r_- \right] \left( 1 - \frac{r_-}{r_0} \right)^{- \frac{1 - 2 \alpha - \alpha^2}{1 + \alpha^2}}}{\sqrt{3} \left( 1 + \alpha^2 \right) Q \sqrt{L^2 + m^2 r_0^2 \left( 1 - \frac{r_-}{r_0} \right)^\frac{2 \left( 1 + \alpha \right) \alpha}{1 + \alpha^2}}} \frac{1}{\sqrt{- \Lambda}}.
    \label{eq:DRN_AdS_large_cosmological_constant_String}
\end{eqnarray}
    The local maximum point $r_0$ in the Einstein frame, obtained from~\eqref{eq:DRN_AdS_large_cosmological_constant_Einstein}, is given by
\begin{equation}
    r_0 = \frac{r_-}{1 + \alpha^2}.
\end{equation}
    Because $r_0 > r_-$, no local maximum point exists in the Einstein frame. In the string frame, determined from~\eqref{eq:DRN_AdS_large_cosmological_constant_String}, the point takes the form
\begin{equation}
    r_0 = \frac{1 - 2 \alpha}{1 + \alpha^2} r_-.
\end{equation}
    Consequently, the bound on chaos in the large negative cosmological constant limit applies exclusively to the string frame, and the bound on chaos~\eqref{eq:boudn_on_chaos} with vanishing angular momentum is reduced to
\begin{equation}
    \left. \Delta_\mathrm{S} \right|_{\Lambda \to - \infty} = - \frac{2 \left( 1 + \alpha - \alpha^2 \right) \left( 1 - 2 \alpha \right)^2 r_-^2}{9 \alpha} \left( \frac{\alpha^2 + 2 \alpha}{2 \alpha - 1} \right)^\frac{4 \alpha^2}{1 + \alpha^2} \Lambda^2 + \mathcal{O}(\Lambda^{-1}).
\end{equation}
    The bound on chaos is negative for
\begin{equation}
     -2 < \alpha < - \frac{3 + \sqrt{17}}{2},
\end{equation}
    yielding
\begin{equation}
     \left. \Delta_\mathrm{S} \right|_{\Lambda \to - \infty} < 0,
\end{equation}
\noindent implying violations of the bound on chaos for the corresponding parameter range in the string frame.

\section{Kerr--Sen--Ads Black Hole} \label{sec_KS}
    In heterotic string theories, the Neveu--Schwarz--Neveu--Schwarz (NS--NS) or the bosonic sector of the four-dimensional effective action in the Einstein frame is given by
\begin{equation}
    S_\mathrm{E} = \int \mathrm{d}^4x \sqrt{-g} \left[ R - \frac{1}{2} \partial_\mu \phi \partial^\mu \phi - e^{-\phi} F_{\mu \nu} F^{\mu \nu} - \frac{e^{- 2 \phi}}{12} H_{\mu \nu \rho} H^{\mu \nu \rho} \right],
\end{equation}
\noindent where $R$ is the Ricci scalar, $\phi$ is the dilaton scalar field, $F_{\mu \nu}$ is the $U(1)$ Maxwell field strength associated with the gauge potential $A_{\mu}$, and $H_{\mu \nu \rho}$ is the field strength of the Kalb--Ramond NS--NS \textit{B}-field, which includes a gauge Chern--Simons term, as follows:
\begin{eqnarray}
    F_{\mu \nu} &=& \partial_\mu A_\nu - \partial_\nu A_\mu,
    \\
    H_{\mu \nu \rho} &=& \partial_\mu B_{\nu \rho} + \partial_\nu B_{\rho \mu} + \partial_\rho B_{\mu \nu} - 2 \left( A_\mu F_{\nu \rho} + A_\nu F_{\rho \mu} + A_\rho F_{\mu \nu} \right).
\end{eqnarray}
    Modifying the definition of the three-form field strength allows the incorporation of a nonzero cosmological constant and the construction of asymptotically AdS solutions via dualization procedure. In this formulation, the three-form field $H_{\mu \nu \rho}$ is expressed in terms of a pseudoscalar field $\chi$, which is often referred to as the axion field in the dual formulation. The modified definition is given by
\begin{equation}
    H = \mathrm{d}B - 2 A \wedge F \equiv - e^{2 \phi} \star \mathrm{d}\chi,
\end{equation}
\noindent where $\star$ denotes the Hodge star operator. The dualization allows for the incorporation of a cosmological constant term into the four-dimensional effective Lagrangian, enabling asymptotically AdS solutions. The resulting Lagrangian in the Einstein frame~\cite{Wu:2020cgf, Ali:2023ppg, Lee:2025vih} takes the form
\begin{equation}
    \mathcal{L}_\mathrm{E} = \sqrt{-g} \left\{ R \! - \! \frac{1}{2} \partial_\mu \phi \partial^\mu \phi \! - \! \frac{e^{2 \phi}}{2} \partial_\mu \chi \partial^\mu \chi \! - \! e^{-\phi} F_{\mu \nu} F^{\mu \nu} \! + \! \chi F_{\mu \nu} \tilde{F}^{\mu \nu} \! - \! \frac{\Lambda}{3} \left[ 4 \! + \! e^{-\phi} \! + \! e^{\phi} \left( 1 \! + \! \chi^2 \right) \right] \right\},
\end{equation}
\noindent where $\tilde{F}^{\mu \nu} = \frac{1}{2} \epsilon^{\mu \nu \rho \sigma} F_{\rho \sigma}$ denotes the dual tensor of $F^{\mu \nu}$; $\epsilon^{\mu \nu \rho \sigma}$ is the Levi-Civita tensor; and $\Lambda$ is the cosmological constant. The Kerr--Sen--AdS metric in the Einstein frame, written in four-dimensional Boyer--Lindquist coordinates, reads
\begin{equation}
    \mathrm{d}s_\mathrm{E}^2 = - \frac{\Delta_r}{\rho^2} \left( \mathrm{d}t - \frac{a \sin^2\theta}{\Xi} \mathrm{d}\varphi \right)^2 + \frac{\rho^2}{\Delta_r} \mathrm{d}r^2 + \frac{\rho^2}{\Delta_\theta} \mathrm{d}\theta^2 + \frac{\Delta_\theta \sin^2\theta}{\rho^2} \left( a dt - \frac{r^2 + 2 b r + a^2}{\Xi} \mathrm{d}\varphi \right)^2,
    \label{eq_metric}
\end{equation}
    where
\begin{eqnarray}
    \rho^2 &=& r^2 + 2 b r + a^2 \cos^2\theta, \qquad \Xi \;\; = \;\; 1 + \frac{\Lambda}{3} a^2, \nonumber
    \\
    \Delta_r &=& \left( r^2 + 2 b r + a^2 \right) \left[ 1 - \frac{\Lambda}{3} \left( r^2 + 2 b r \right) \right] - 2 M r, \nonumber
    \\
    \Delta_\theta &=& 1 + \frac{\Lambda}{3} a^2 \cos^2\theta, \nonumber
\end{eqnarray}
\noindent and the dilaton field $\phi$ and pseudoscalar field $\chi$ are obtained as
\begin{equation}
    \phi(r) = - \ln \left( \frac{\rho^2}{r^2 + a^2 \cos^2\theta} \right), \quad \chi(r) = \frac{2 b a \cos\theta}{r^2 + a^2 \cos^2\theta}.
\end{equation}
    The Kerr--Sen--AdS metric is characterized by the mass $M$, angular momentum $J$, and electric charge $Q$. The spin parameter is defined as $a = J/M$, and the dilatonic scalar charge as $b = Q^2 / ( 2 M )$. In the absence of a cosmological constant $(\Lambda = 0)$ and vanishing spin parameter $(a = 0)$, the metric is reduced to the electrically charged GMGHS solution~\cite{Gibbons:1987ps, Garfinkle:1990qj} via a radial coordinate shift. The electromagnetic potential, assuming the absence of magnetic charge, is given by
\begin{equation}
    A = - \frac{Q r}{\rho^2} \left( \mathrm{d}t - \frac{a \sin^2\theta}{\Xi} \mathrm{d}\varphi \right).
\end{equation}
    
    In the Kerr--Sen--AdS background~\eqref{eq_metric}, the angular velocity of an observer at spatial infinity is defined by
\begin{equation}
    \Omega_\infty = \left. - \frac{g_{t\varphi}}{g_{\varphi\varphi}} \right|_{r \to \infty} = \frac{a \Lambda}{3}.
\end{equation}
    We define the frame of an asymptotically static observer by performing a coordinate transformation~\cite{Hawking:1998kw,Gwak:2018akg,Gwak:2021tcl}, eliminating the angular velocity at spatial infinity:
\begin{equation}
    \varphi \to \varphi + \frac{a \Lambda}{3} t.
\end{equation}
    After coordinate transformation, the Kerr--Sen--AdS metric is reduced to
\begin{eqnarray}
    \mathrm{d}s_\mathrm{E}^2 &=& - \frac{\Delta_r}{\Xi^2 \rho^2} \left( \Delta_\theta \mathrm{d}t - a \sin^2\theta \mathrm{d}\varphi \right)^2 + \frac{\rho^2}{\Delta_r} \mathrm{d}r^2 + \frac{\rho^2}{\Delta_\theta} \mathrm{d}\theta^2 \nonumber
    \\
    && \quad + \frac{\Delta_\theta \sin^2\theta}{\Xi^2 \rho^2} \left\{ a \left[ 1 - \frac{\Lambda}{3} \left( r^2 + 2 b r \right) \right] \mathrm{d}t - \left( r^2 + 2 b r + a^2 \right) \mathrm{d}\varphi \right\}^2
\end{eqnarray}
\noindent and the corresponding potential is represented as
\begin{equation}
    A =  - \frac{Q r}{\Xi \rho^2} \left( \Delta_\theta \mathrm{d}t - a \sin^2\theta \mathrm{d}\varphi \right).
\end{equation}
    The Hawking temperature is invariant under the transformation and takes the form
\begin{equation}
    T_\mathrm{H} = \frac{r_\mathrm{h} \left[ 1 - \frac{a^2}{r_\mathrm{h}^2} - \frac{\Lambda}{3} a^2 - \frac{\Lambda}{3} \left( r_\mathrm{h} + 2 b \right) \left( 3 r_\mathrm{h} + 2 b \right) \right]}{4 \pi \left( r_\mathrm{h}^2 + 2 b r_\mathrm{h} + a^2 \right)},
\end{equation}
\noindent where $r_\mathrm{h}$ denotes the outer horizon of the Kerr--Sen--AdS black hole. Correspondingly, the surface gravity is determined by
\begin{equation}
    \kappa = 2 \pi T_\mathrm{H} = \frac{r_\mathrm{h} \left[ 1 - \frac{a^2}{r_\mathrm{h}^2} - \frac{\Lambda}{3} a^2 - \frac{\Lambda}{3} \left( r_\mathrm{h} + 2 b \right) \left( 3 r_\mathrm{h} + 2 b \right) \right]}{2 \left( r_\mathrm{h}^2 + 2 b r_\mathrm{h} + a^2 \right)}.
    \label{eq_KS-AdS_surface_gravity}
\end{equation}

\subsection{Asymptotically Flat Spacetime}
    In asymptotically flat spacetime $( \Lambda = 0 )$, the Kerr--Sen black hole possesses an outer horizon $r_\mathrm{h}$, where $r_\mathrm{h} = r_+$, and inner Cauchy horizon $r_-$, which are given by
\begin{equation}
    r_\pm = M - b \pm \sqrt{\left( M - b \right)^2 - a^2}.
\end{equation}
    The parameters are constrained by
\begin{equation}
    M \geq \left| a \right| + b.
\end{equation}
    For simplicity, the spin parameter $a$ is assumed to be positive,  and $|a|$ is replaced with $a$. The charge of the black hole, $Q$, is fixed as positive, whereas the sign of the charge of the particle, $q$, is varied to investigate both attractive and repulsive couplings.

\subsubsection{Extremal Case}
    The extremal Kerr--Sen black hole is characterized by the coincidence of the outer and inner horizons, with the outer horizon $r_\mathrm{h}$ expressed as
\begin{equation}
    r_\mathrm{h} = a,
\end{equation}
\noindent and the mass attains the extremal value
\begin{equation}
    M = a + b.
\end{equation}
    The surface gravity~\eqref{eq_KS-AdS_surface_gravity} vanishes in both the Einstein and string frames:
\begin{equation}
    \left. \kappa_\mathrm{E} \right|_{\Lambda = 0} = \left. \kappa_\mathrm{S} \right|_{\Lambda = 0} = 0.
\end{equation}

    \noindent \textbf{\textit{Vanishing angular momentum in near-horizon limit}}---The case of vanishing angular momentum in the near-horizon region is considered. The derivatives of the effective potential in both frames are reduce to
\begin{eqnarray}
    \left. V'_\mathrm{E}(r_0) \right|_{\{ \Lambda = 0, \, L = 0, \, r_0 = r_\mathrm{h} + \epsilon \}} &=& \frac{1}{4 \left( a + b \right)^2} \sqrt{q^2 Q^2 + 4 m^2 \left( a + b \right)^2 \left( 1 + \frac{2 b}{a} \right)} + \mathcal{O}(\epsilon),
    \\
    \left. V'_\mathrm{S}(r_0) \right|_{\{ \Lambda = 0, \, L = 0, \, r_0 = r_\mathrm{h} + \epsilon \}} &=& \frac{1}{4 \left( a + b \right)^2} \sqrt{q^2 Q^2 + 4 m^2 \left( a + b \right)^2} + \mathcal{O}(\epsilon).
\end{eqnarray}
    Because the derivative of the effective potential remains positive, no local maximum exists in the near-horizon region with vanishing angular momentum. Consequently, the conditions required for the definition of the Lyapunov exponent~\eqref{eq:reveiw_lyapunov_exponent} are not satisfied, and the corresponding computation is omitted.

    \noindent \textbf{\textit{Near-horizon limit}}---The near-horizon limit of the extremal Kerr--Sen black hole requires nonzero angular momentum for the existence of a local maximum, as indicated by the findings of the preceding analysis. In the near-horizon limit, the charge of the particle, $q$, in both frames is determined by
\begin{eqnarray}
    \left. q_\mathrm{E}^\pm \right|_{\{ \Lambda = 0, \, r_0 = r_\mathrm{h} + \epsilon \}} &=& \frac{L \left( a + 2 b \right) \pm 2 \left( a + b \right) \sqrt{L^2 - m^2 a \left( a + 2 b \right)}}{Q a},
    \\
    \left. q_\mathrm{S}^\pm \right|_{\{ \Lambda = 0, \, r_0 = r_\mathrm{h} + \epsilon \}} &=& \frac{L \left( a + 2 b \right) \pm 2 \left( a + b \right) \sqrt{L^2 - m^2 a^2}}{Q a}.
\end{eqnarray}
    In contrast to the case of vanishing angular momentum, where no local maximum exists, the presence of nonzero angular momentum allows the analysis of a local maximum. Within the near-horizon limit with nonzero angular momentum, the bound on chaos~\eqref{eq:boudn_on_chaos} in both frames is reduced to
\begin{eqnarray}
    \left. \Delta_\mathrm{E}^2 \right|_{\{ \Lambda = 0, \, r_0 = r_\mathrm{h} + \epsilon \}} \!\!\!\!\! &=& \!\!\!\!\! \frac{- 2 L^2 + 2 m^2 a \left( a + b \right) \pm \frac{L a + \left| L \right| \left( a + 2 b \right)}{a + b} \sqrt{L^2 - m^2 a \left( a + 2 b \right)}}{4 L^2 a^3 \left( a + b \right)^2} \epsilon^3 + \mathcal{O}(\epsilon^4),
    \label{eq:Einstein_KS_extreme_near_horizon}
    \\
    \left. \Delta_\mathrm{S}^2 \right|_{\{ \Lambda = 0, \, r_0 = r_\mathrm{h} + \epsilon \}} \!\!\!\!\! &=& \!\!\!\!\! \frac{- 2 L^2 + 2 m^2 a^2 \pm \frac{L a + \left| L \right| \left( a + 2 b \right)}{a + b} \sqrt{L^2 - m^2 a^2}}{4 L^2 a^3 \left( a + b \right)^2} \epsilon^3 + \mathcal{O}(\epsilon^4),
    \label{eq:string_KS_extreme_near_horizon}
\end{eqnarray}
\noindent where the upper plus and lower minus signs in~\eqref{eq:Einstein_KS_extreme_near_horizon}~and~\eqref{eq:string_KS_extreme_near_horizon} correspond to the charges of the particle, $q^\pm$, respectively. For $q = q^+$, the violation of the bound on chaos in the Einstein frame occurs when
\begin{equation}
    \frac{L}{m} < - \sqrt{a^2 + 2 a b} \quad \text{or} \quad \sqrt{a^2 + 2 a b} < \frac{L}{m} < a + b,
\end{equation}
\noindent whereas in the string frame, the violation is constrained by
\begin{equation}
    \frac{L}{m} < - \frac{\sqrt{a^3 + 2 a^2 b + a b^2}}{a + 2 b}.
\end{equation}
    For $q = q^-$, in the Einstein frame, the violation occurs when
\begin{equation}
    \frac{L^2}{m^2} > a^2 + 2 a b,
\end{equation}
\noindent whereas in the string frame, the condition for violation is 
\begin{equation}
    \frac{L^2}{m^2} > a^2.
\end{equation}

    \noindent \textbf{\textit{Large angular momentum}}---Considering the large angular momentum limit, the electric charge of the particle, $q$, coincides in both frames and is given by
\begin{equation}
\begin{aligned}
    \frac{q}{L} =
    \begin{cases}
        - \frac{1}{\left( r_0 + a + 2 b \right)^2} & \left( L \gg 0 \right)
        \\
        \frac{r_0^2 - 4 a r_0 - 4 a b}{\left[ r_0^2 - \left( a - 2 b \right) r_0 + 2 a^2 \right]^2} & \left( L \ll 0 \right)
    \end{cases}.
\end{aligned}
\end{equation}
    At the large angular momentum limit, the effective potential exhibits no local maximum for positive angular momentum, whereas a local maximum emerges when the angular momentum is negative. In the latter case, the local maximum point $r_0$ is defined by
\begin{equation}
    r_0 = 2 a \left( 1 + \sqrt{1 + \frac{b}{a}} \right).
\end{equation}
    The bound on chaos~\eqref{eq:boudn_on_chaos} in both frames is identically reduced to
\begin{equation}
    \left. \Delta_\mathrm{E}^2 \right|_{\{ \Lambda = 0, \, L \to - \infty \}} = \left. \Delta_\mathrm{S}^2 \right|_{\{ \Lambda = 0, \, L \to - \infty \}} = - \frac{8 a \left( a + b \right) \left( 2 r_0 - a + 2 b \right) \left( r_0 - a \right)^3}{r_0 \left( r_0 + 2 b \right)\left[ \left( 3 a + 2 b \right) r_0 + 2 a \left( a + 2 b \right) \right]^3} + \mathcal{O}(L^{-1}) < 0,
\end{equation}
\noindent indicating that the bound on chaos is violated at the large negative angular momentum limit.

\subsubsection{Non-Extremal Case}
    The analysis considers a Kerr--Sen black hole with non-extremal parameters, where the outer and inner horizons are distinct. The outer horizon $r_\mathrm{h}$ is located at
\begin{equation}
    r_\mathrm{h} = M - b + \sqrt{\left( M - b \right)^2 - a^2},
\end{equation}
\noindent and the surface gravity~\eqref{eq_KS-AdS_surface_gravity} in both frames is reduced to
\begin{equation}
    \left. \kappa_\mathrm{E} \right|_{\Lambda = 0} = \left. \kappa_\mathrm{S} \right|_{\Lambda = 0} = \frac{r_\mathrm{h}^2 - a^2}{2 r_\mathrm{h} \left( r_\mathrm{h}^2 + 2 b r_\mathrm{h} + a^2 \right)}.
\end{equation}

    \noindent \textbf{\textit{Near-horizon limit}}---The near-horizon limit is considered for a particle with nonzero angular momentum. The charge of the particle~\eqref{eq:q} in both frames takes the form
\begin{eqnarray}
    \left. q_\mathrm{E}^\pm \right|_{\{ \Lambda = 0, \, r_0 = r_\mathrm{h} + \epsilon \}} &=& \frac{L \left( r_\mathrm{h} + 2 b \right) \pm i m \left( r_\mathrm{h}^2 + 2 b r_\mathrm{h} + a^2 \right) \sqrt{1 + \frac{2 b}{r_\mathrm{h}}}}{Q a},
    \label{eq:KS_NH_Einstein}
    \\
    \left. q_\mathrm{S}^\pm \right|_{\{ \Lambda = 0, \, r_0 = r_\mathrm{h} + \epsilon \}} &=& \frac{L \left( r_\mathrm{h} + 2 b \right) \pm i m \left( r_\mathrm{h}^2 + 2 b r_\mathrm{h} + a^2 \right)}{Q a}.
    \label{eq:KS_NH_String}
\end{eqnarray}
    Both expressions~\eqref{eq:KS_NH_Einstein}~and~\eqref{eq:KS_NH_String} are complex, thus precluding the existence of a local maximum of the effective potential for any real value of the charge.

\begin{figure}[H]
    \centering
    \includegraphics[width=14.0cm]{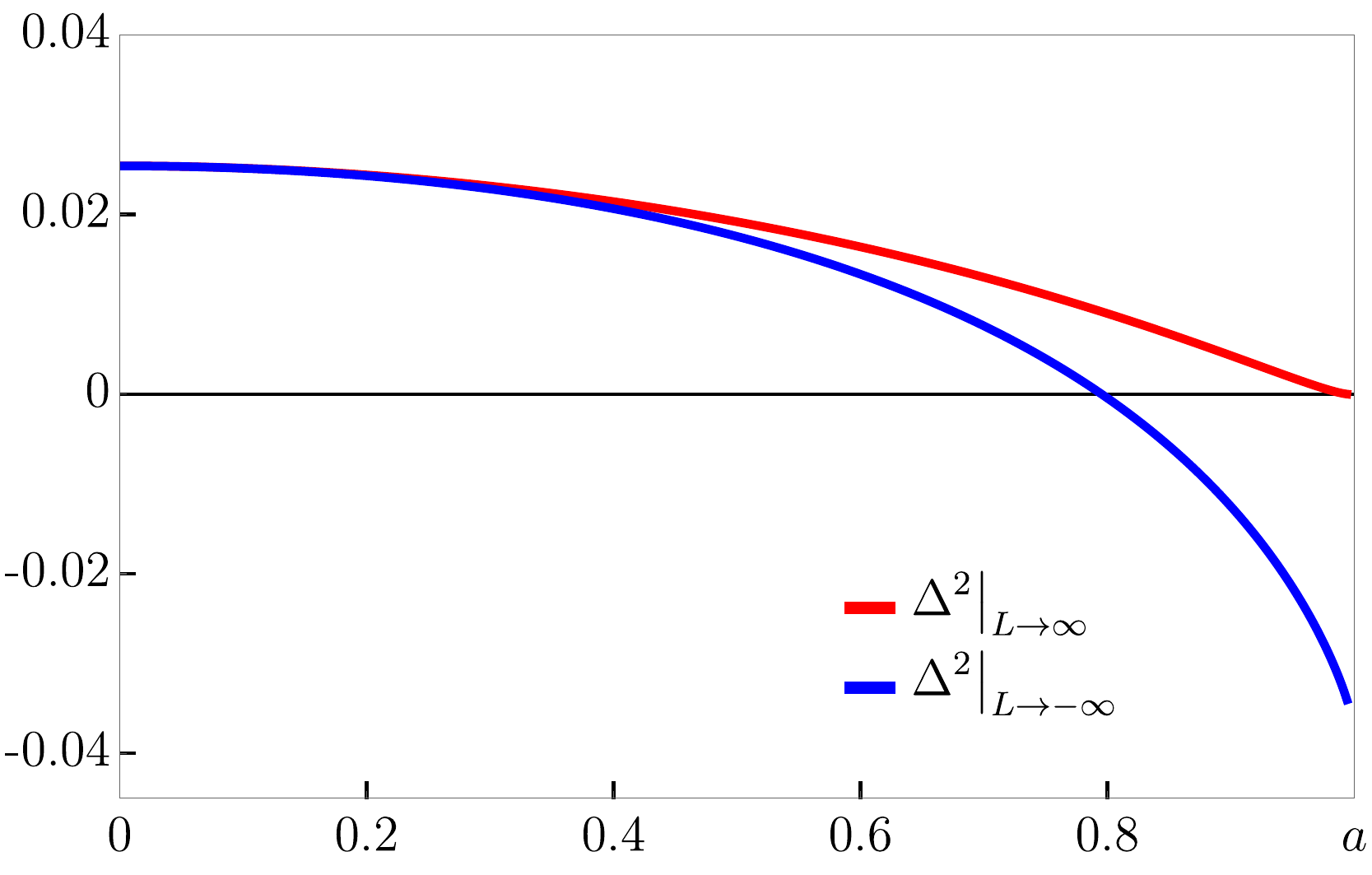}
    \caption{Bound on chaos in the large angular momentum limit of non-extremal Kerr--Sen black hole backgrounds.}
    \label{fig:KS_LAM}
\end{figure}

    \noindent \textbf{\textit{Large angular momentum}}---In the limit of large angular momentum, the mass of a black hole is determined in both frames by
\begin{equation}
    M = \frac{\left( r_0 + b \right) \left\{ \left( r_0 + 2 b \right) \left( 3 r_0 + 2 b \right) + 2 a \left[ a \pm \sqrt{a^2 + \left( r_0 + 2 b \right) \left( 3 r_0 + 2 b \right)} \right] \right\}}{\left( 3 r_0 + 2 b^2 \right)^2},
\end{equation}
\noindent where the upper plus and lower minus signs correspond to the large positive angular momentum and negative angular momentum, respectively. In both frames, the bounds on chaos~\eqref{eq:boudn_on_chaos} reduce to an identical form in the limits of large positive and negative angular momenta, respectively, and are expressed as
\begin{eqnarray}
    && \left. \Delta_\mathrm{E}^2 \right|_{\{ \Lambda = 0, \, L \to \infty \}} \;\; = \left. \Delta_\mathrm{S}^2 \right|_{\{ \Lambda = 0, \, L \to \infty \}} \:\:\, = \, \mathcal{O}(1),
    \label{eq:KS_positive_large_angular_momentum}
    \\
    && \left. \Delta_\mathrm{E}^2 \right|_{\{ \Lambda = 0, \, L \to - \infty \}} = \left. \Delta_\mathrm{S}^2 \right|_{\{ \Lambda = 0, \, L \to - \infty \}} = \, \mathcal{O}(1).
    \label{eq:KS_negative_large_angular_momentum}
\end{eqnarray}
    The complex structure of the analytic expressions~\eqref{eq:KS_positive_large_angular_momentum}~and~\eqref{eq:KS_negative_large_angular_momentum} is numerically depicted in Fig.~\ref{fig:KS_LAM}.
    Fig.~\ref{fig:KS_LAM} depicts the bound on chaos in the large angular momentum limit of Kerr--Sen black holes in the Einstein and string frames, corresponding to~\eqref{eq:KS_positive_large_angular_momentum}~and~\eqref{eq:KS_negative_large_angular_momentum}. The parameters are set to $M = 1$, $Q = 0.1$, and $a < 0.995$ to ensure a non-extremal black hole. The horizontal axis represents $a$, whereas the vertical axis represents the bound on chaos, $\Delta^2$. The red line indicates the bound in the limit of large positive angular momentum, whereas the blue line indicates the bound in the limit of large negative angular momentum. For positive angular momentum, the bound remains positive within the range of $a$ shown in Fig.~\ref{fig:KS_LAM}, whereas for negative angular momentum, the bound changes to negative when $a > 0.7954$. Violations occur when the signs of $a$ and $L$ are opposite and $a$ exceeds a critical value.

\subsection{Asymptotically AdS Spacetime}
    The Kerr--Sen black hole in an asymptotically AdS background, characterized by a negative cosmological constant $(\Lambda < 0)$, has an outer horizon $r_\mathrm{h}$ that satisfies
\begin{equation}
    r_- < r_\mathrm{h} < r_+,
\end{equation}
\noindent where, in contrast to the asymptotically flat case, $r_+$ and $r_-$ are not identified as the outer and inner horizons, respectively.

\subsubsection{Extremal Case}
    In the extremal configuration, the black hole mass $M$ and cosmological constant $\Lambda$ in both frames are given by
\begin{equation}
    M = \frac{1}{2} \left( r_\mathrm{h}^2 + 2 b r_\mathrm{h} + a^2 \right) \left[ 1 - \frac{\Lambda}{3} \left( r_\mathrm{h} + 2 b \right) \right], \quad \Lambda = - \frac{3 \left( a^2 - r_\mathrm{h}^2 \right)}{\left[ a^2 + \left( r_\mathrm{h} + 2 b \right) \left( 3 r_\mathrm{h} + 2 b \right) \right] r_\mathrm{h}^2}.
\end{equation}
    For the cosmological constant to describe an AdS background $( \Lambda < 0 )$, the condition
\begin{equation}
    r_\mathrm{h} < a
\end{equation}
\noindent is required. The surface gravity in both frames vanishes identically:
\begin{equation}
    \kappa_\mathrm{E} = \kappa_\mathrm{S} = 0.
\end{equation}

\begin{figure}[H]
    \centering
    \includegraphics[width=14.0cm]{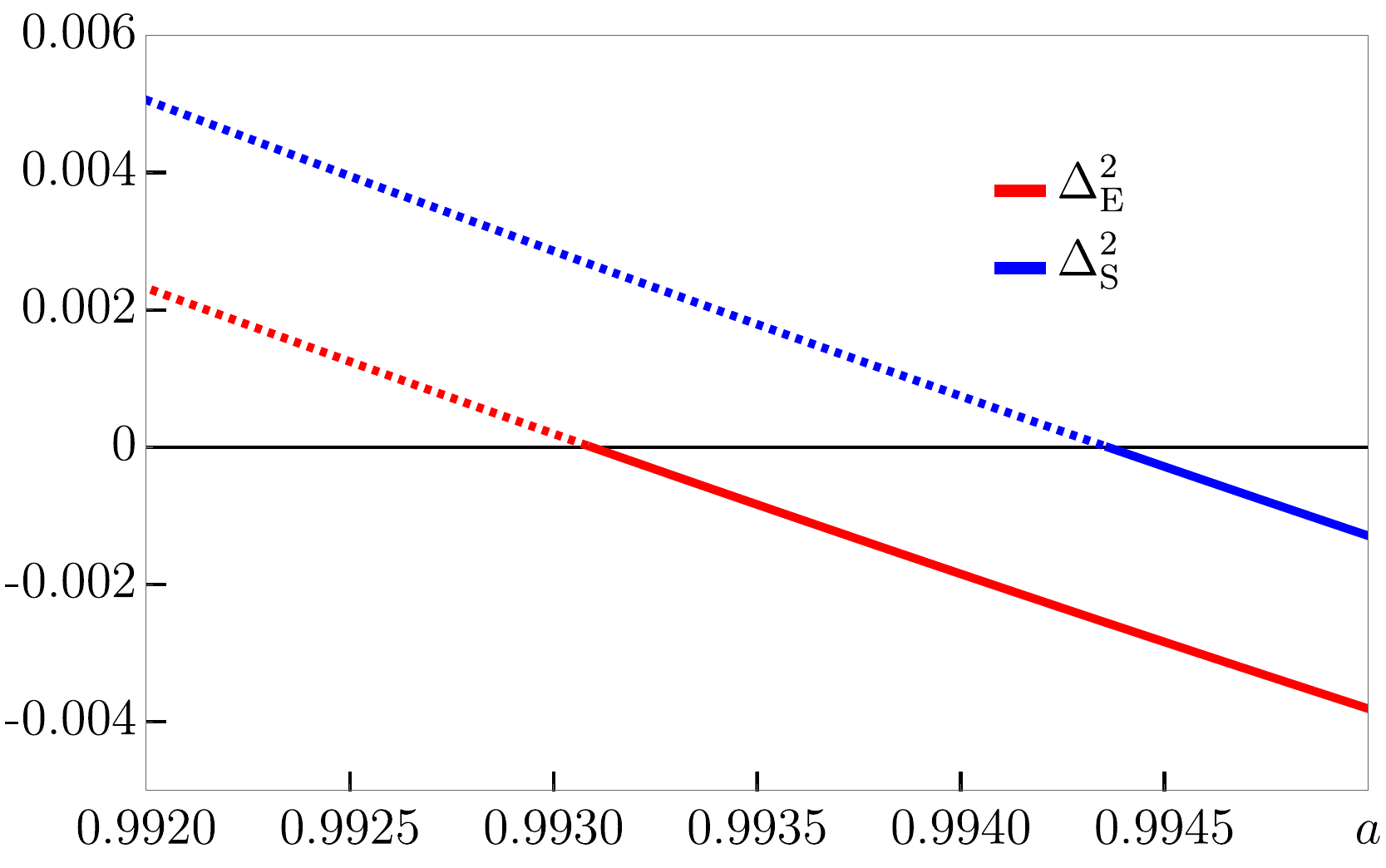}
    \caption{Bound on chaos in the near-horizon limit with vanishing angular momentum for extremal Kerr--Sen--AdS black hole backgrounds.}
    \label{fig:KS_AdS_Ext_NH}
\end{figure}

    \noindent \textbf{\textit{Vanishing angular momentum in near-horizon limit}}---We consider the case of a vanishing angular momentum. The bound on chaos~\eqref{eq:boudn_on_chaos} in both frames is reduced to
\begin{equation}
    \left. \Delta_\mathrm{E} \right|_{\{ L = 0, \, r_0 = r_\mathrm{h} + \epsilon \}} = \mathcal{O}(\epsilon^3), \quad \left. \Delta_\mathrm{S} \right|_{\{ L = 0, \, r_0 = r_\mathrm{h} + \epsilon \}} = \mathcal{O}(\epsilon^3).
    \label{eq:KS_AdS_extreme_near_horizon}
\end{equation}
    The expression~\eqref{eq:KS_AdS_extreme_near_horizon} is of considerable length, and thus, the results are shown numerically in Fig.~\ref{fig:KS_AdS_Ext_NH}.
    Fig.~\ref{fig:KS_AdS_Ext_NH} shows the bound on chaos in the near-horizon region of extremal Kerr--Sen--AdS black holes with vanishing angular momentum in both the Einstein and string frames, corresponding to~\eqref{eq:KS_AdS_extreme_near_horizon}. The parameters are set to $M = 1$, $Q = 0.1$, $m = 1$, $q = 0.1$, $\alpha < 0.995$, and $\epsilon = 1$ for simplicity. The horizontal axis represents the spin parameter $a$, whereas the vertical axis represents the bound on chaos, $\Delta^2$. The red line indicates the bound in the Einstein frame, whereas the blue line indicates the bound in the string frame. Solid lines signify negative values of the bound on chaos, whereas dashed lines signify positive values. The bound in the Einstein frame is negative for $\alpha > 0.9931$. By comparison, the bound in the string frame is negative for all $\alpha > 0.9944$. Similarly, as shown in Fig.~\ref{fig:DRN_AdS_Ext_NH}, the positive region of the bound on chaos remains inconsistent with the Lyapunov exponent evaluated at the local maximum and constitutes an artifact of the analytic expressions in~\eqref{eq:KS_AdS_extreme_near_horizon}.

\begin{figure}[H]
    \centering
    \includegraphics[width=14.0cm]{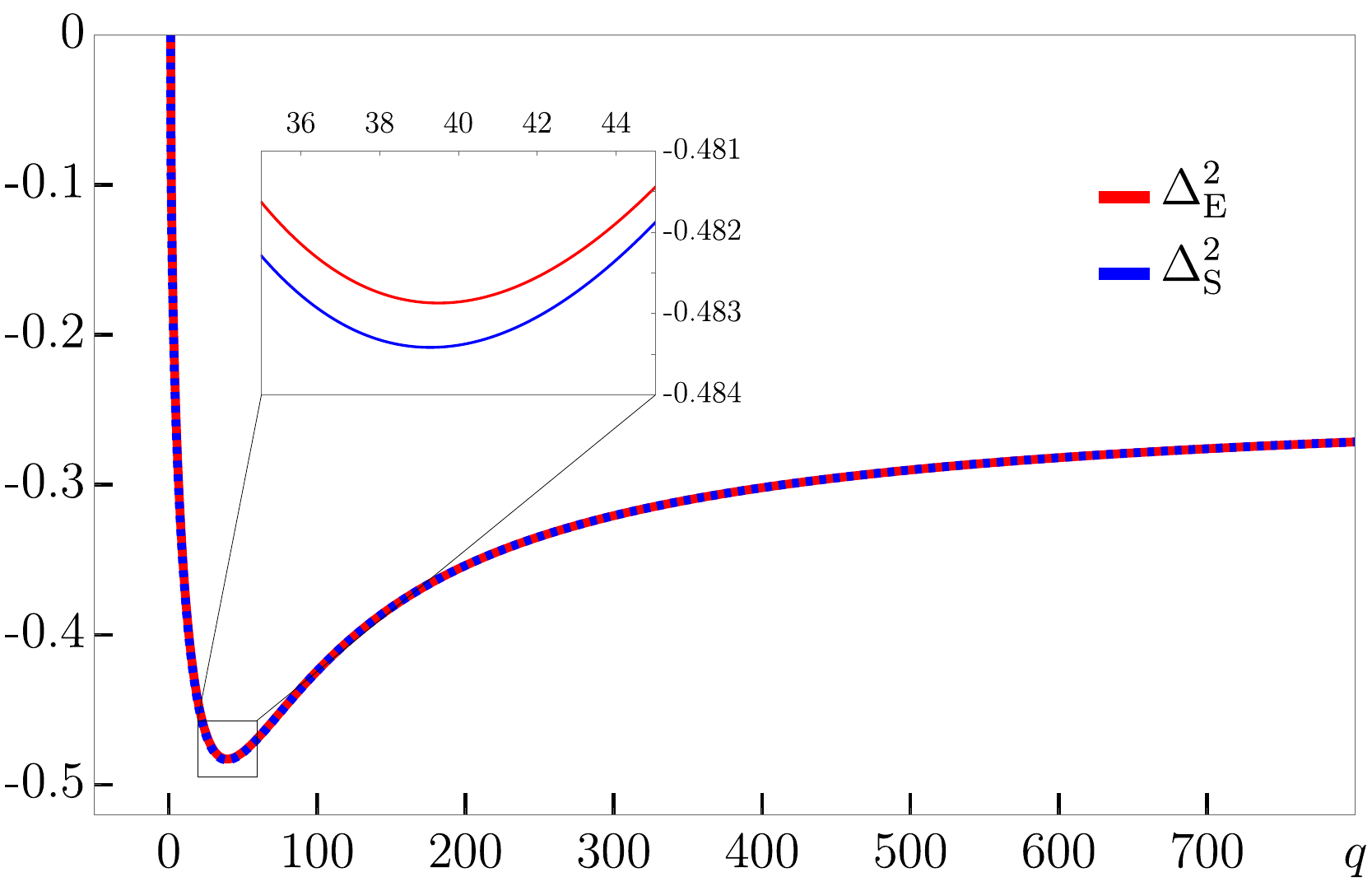}
    \caption{Bound on chaos in the extremal Kerr--Sen--AdS black hole background.}
    \label{fig:KS_AdS_Ext}
\end{figure}

    \noindent \textbf{\textit{General Case}}---In the following analysis, the bound on chaos for extremal Kerr--Sen--AdS black holes is evaluated without imposing particular limiting conditions. The results are presented numerically in Fig.~\ref{fig:KS_AdS_Ext}.
    Fig.~\ref{fig:KS_AdS_Ext} illustrates the bound on chaos in an extremal Kerr--Sen--AdS black hole. The parameters are set to $M = 1$, $Q = 0.908$, $\Lambda = -0.5$, $a = 0.5$, $m = 1$, and $L = -20$. The horizontal axis represents the electric charge of the particle, $q$, whereas the vertical axis represents the bound on chaos, $\Delta^2$. The red line indicates the bound in the Einstein frame, whereas the blue line indicates the bound in the string frame. The extremal condition ensures a negative bound on chaos across the entire range of $q$. The results for both frames are nearly identical, whereas the bound on chaos begins to be evaluated from $q = 1.4535$ in the Einstein frame, and from $q = 1.2679$ in the string frame. The magnified view shown in Fig.~\ref{fig:KS_AdS_Ext} shows that the two frames remain distinguishable with a difference of $5 \times 10^{-4}$ at $q = 40$.

\subsubsection{Non-Extremal Case}
    We consider the Kerr--Sen--AdS black hole in the non-extremal regime. In contrast to the extremal case, in which the horizons coincide, the outer and inner horizons are distinct. The behavior of the effective potential depends sensitively on the value of the cosmological constant.
    
    \noindent \textbf{\textit{Large negative cosmological constant}}---In the limit of the large negative cosmological constant $(\Lambda \to - \infty)$, the effective potential in both frames is given by
\begin{equation}
    \left. V_\mathrm{E}(r) \right|_{\Lambda \to - \infty} = \left. V_\mathrm{S}(r) \right|_{\Lambda \to - \infty} = - \sqrt{\frac{L^2 \left( r^2 + 2 b r \right)}{3 \left( r^2 + 2 b r + a^2 \right)}} \sqrt{-\Lambda} + \mathcal{O}\left( \frac{1}{\sqrt{-\Lambda}} \right).
\end{equation}
    The effective potential in both frames decreases monotonically as the cosmological constant decreases further into negative values, without forming a local maximum. The absence of a local maximum prevents the existence of unstable circular orbits and precludes the definition of the Lyapunov exponent. This expression is consistent with the results in~\cite{Lee:2025vih}.

\begin{figure}[H]
    \centering
    \includegraphics[width=14.0cm]{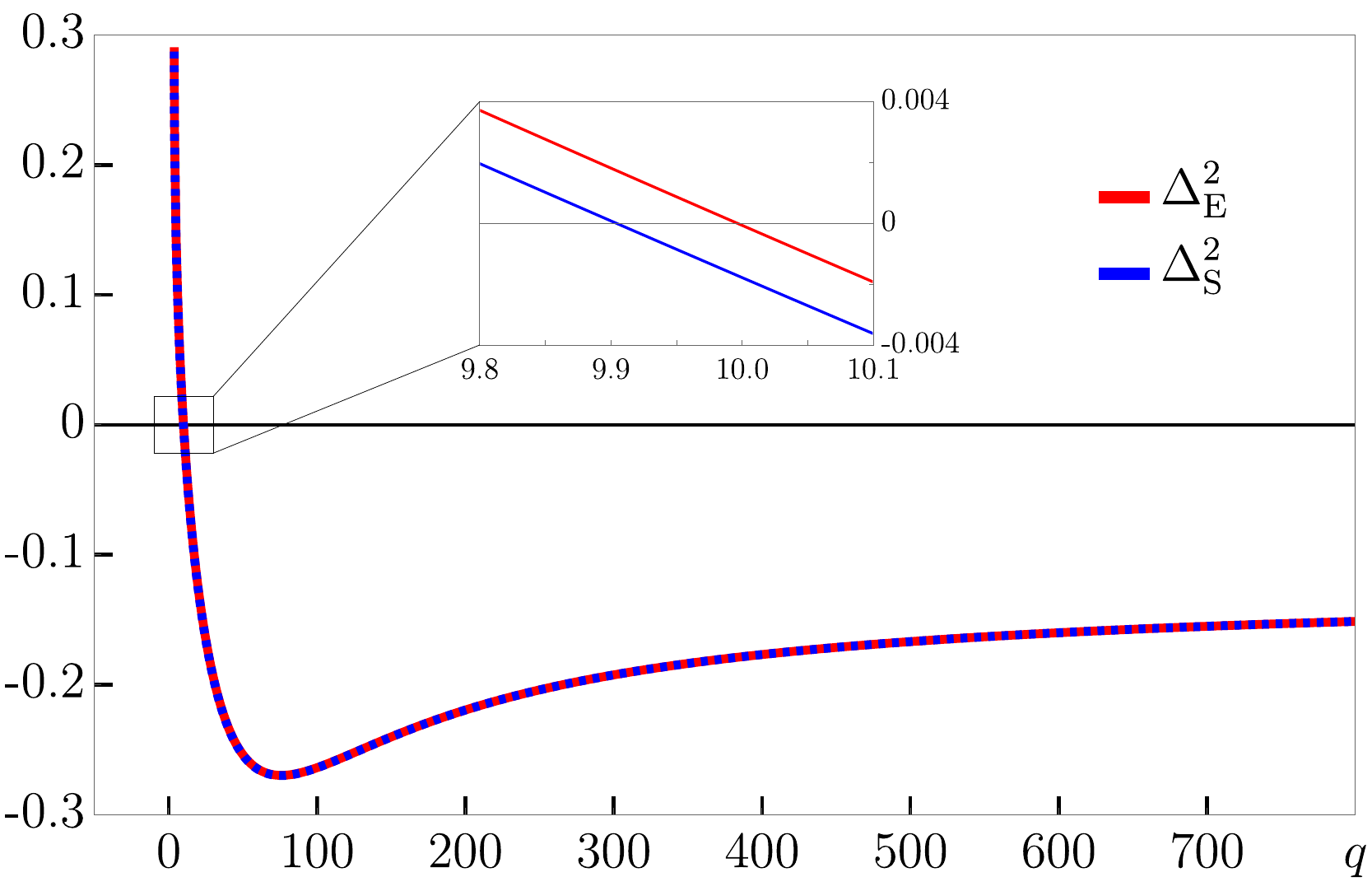}
    \caption{Bound on chaos in the non-extremal Kerr--Sen--AdS black hole background.}
    \label{fig:KS_AdS_NonExt}
\end{figure}

    \noindent \textbf{\textit{General Case}}---Given the complexity of the expression for the bound on chaos of non-extremal Kerr--Sen--AdS black holes, numerical evaluation is depicted in Fig.~\ref{fig:KS_AdS_NonExt} without applying particular limits.
    Fig.~\ref{fig:KS_AdS_NonExt} depicts the bound on chaos in a non-extremal Kerr--Sen--AdS black hole. The parameters are set to $M = 1$, $Q = 0.454$, $\Lambda = -0.5$, $a = 0.5$, $m = 1$, and $L = -20$. The horizontal axis represents the electric charge of the particle, $q$, whereas the vertical axis represents the bound on chaos, $\Delta^2$. The red line indicates the bound in the Einstein frame, whereas the blue line indicates the bound in the string frame. The bounds in both frames are virtually identical, although the evaluation begins at $q = 3.7790$ in the Einstein frame and at $q = 3.6914$ in the string frame. The magnified view shown in Fig.~\ref{fig:KS_AdS_NonExt} reveals that the bound turns negative for $q > 9.996$ in the Einstein frame and for $q > 9.904$ in the string frame. Thus, within the interval $9.904 < q < 9.996$, the bound on chaos holds in the Einstein frame, whereas it is violated in the string frame. The discrepancy between the two frames amounts to $2 \times 10^{-3}$ at $q = 10$.

\section{Conclusion} \label{sec_conclusion}
    In this study, we investigated the frame dependence of the Lyapunov exponent and the bound on chaos in dilatonic RN--AdS and Kerr--Sen--AdS black holes. The analysis is motivated by the consideration of conformal transformations between the Einstein and string frames, which could potentially alter the representation of physical quantities and raise the question of universality or frame dependence of chaotic behavior. We analyzed the dynamics near unstable orbits and computed the corresponding local Lyapunov exponents in the Einstein and string frames. The Lyapunov exponent is compared with the surface gravity of the black hole, which represents the upper bound on chaos proposed in two referenced studies~\cite{Maldacena:2015waa, Hashimoto:2016dfz}. Comparison between the Einstein and string frames reveals several aspects of chaotic behavior in string-inspired AdS spacetimes.

    We derived the Lyapunov exponent for a charged particle by constructing the effective Lagrangian in a stationary, axisymmetric, and circular spacetime metric. The resulting expression reveals a general difference in the Lyapunov exponent between the Einstein and string frames, which originates from the coupling between the dilaton field and the metric. At the limit of a vanishing dilaton field, the two frames are trivially equivalent, yielding identical Lyapunov exponents. Analogously, for a massless particle, the dilaton contribution vanishes, and the Lyapunov exponent remains invariant under conformal transformations between the frames. The mass of the particle functions as an effective carrier of the dilaton contribution, which leads to frame dependence in the dynamical behavior. To investigate frame dependence in the bound on chaos within an AdS spacetime background, analysis is performed for dilatonic RN--AdS and Kerr--Sen--AdS black holes, which provide representative cases of the geometries. Each case is analyzed by classifying the spacetime as asymptotically flat or asymptotically AdS, and further by distinguishing between extremal and non-extremal geometries to examine the resulting dynamical behavior.

    In the extremal dilatonic RN black hole, violations of the bound on chaos occur in the Einstein frame for vanishing angular momentum over most regions of the coupling constant $\alpha$, whereas in the string frame, violations appear exclusively for negative $\alpha$. When the angular momentum is large, both frames yield identical values of the bound on chaos, and violations consistently occur. In the near-horizon limit, the values of the bound on chaos differ between the two frames, whereas violations occur in both frames in all cases with $\alpha^2 < 1$. In the non-extremal dilatonic RN black hole, the Einstein frame satisfies the bound on chaos for vanishing angular momentum and in the near-horizon limit. Conversely, the string frame can exhibit violations for negative $\alpha$. In the extremal dilatonic RN--AdS black hole, both frames manifest violations under vanishing angular momentum and near-horizon conditions. The Einstein frame features symmetric violations with respect to positive and negative $\alpha$ values, whereas the string frame yields violations for a single sign of $\alpha$. For large angular momentum, both frames produce identical values of the bound on chaos, accompanied by violations. In the non-extremal dilatonic RN--AdS black hole, the Einstein frame satisfies the bound on chaos for vanishing angular momentum and in the near-horizon limit, whereas the string frame can present violations for negative $\alpha$. Under vanishing angular momentum and in the regime of a large cosmological constant, the Einstein frame is not defined for the bound on chaos because of the absence of a local maximum, whereas the string frame exhibits violations confined to a specific negative interval of $\alpha$.

    In the extremal Kerr--Sen black hole, the absence of a local maximum precludes the evaluation of the bound on chaos for vanishing angular momentum and in the near-horizon limit. Within the near-horizon limit, both frames exhibit violations under distinct parameter ranges. For large angular momentum, positive angular momentum yields no local maximum, whereas negative angular momentum allows for the evaluation of the bound on chaos. In the latter case, both frames yield identical values of the bound on chaos with violations. In the non-extremal Kerr--Sen black hole, the absence of a local maximum in the near-horizon limit implies that no violations occur in either frame. For large angular momentum, positive and negative values of angular momentum lead to different magnitudes of the bound on chaos in the large angular momentum limit, whereas the results remain frame-independent. In the extremal Kerr--Sen--AdS black hole, frame dependence of the bound on chaos appears for vanishing angular momentum and under the near-horizon condition, and the difference persists numerically in the general case. In the non-extremal Kerr--Sen--AdS black hole, both frames present identical effective potentials for large angular momentum. Although the absence of a local maximum precludes evaluation of the bound on chaos in the extremal limit, the difference between the two frames is observed numerically in the general case. The overall results are summarized in Tables~\ref{tab:DRN}~and~\ref{tab:KS}.
\begin{table}[H]
    \centering \renewcommand{\arraystretch}{1.5}
    \begin{tabular}{|c|c|c|c|c|} \hline
        \multicolumn{3}{|c|}{\textbf{Dilatonic RN Black Hole}} & \textbf{Einstein Frame} & \textbf{String Frame}
        \\ \hline
        \multirow{4}{*}{$\mathbf{\Lambda = 0}$} & \multirow{3}{*}{Extremal} & $L = 0$ & $0 < \alpha^2 < 1$ & $-1 < \alpha < 0$
        \\ \cline{3-5}
        & & $L \to \pm \infty$ & Violation & Violation
        \\ \cline{3-5}
        & & $r_0 = r_\mathrm{h} + \epsilon$ & Violation & Violation
        \\ \cline{2-5}
        & Non-Extremal & \makecell{$L = 0$, \\ $r_0 = r_\mathrm{h} + \epsilon$} & \textbf{---} & $\alpha < 0$
        \\ \hline
        \multirow{4}{*}{$\mathbf{\Lambda < 0}$} & \multirow{2}{*}{Extremal} & \makecell{$L = 0$, \\ $r_0 = r_\mathrm{h} + \epsilon$} & $\alpha^2 > 0.1231$ (Fig.~\ref{fig:DRN_AdS_Ext_NH}) & $\alpha < -0.0553$ (Fig.~\ref{fig:DRN_AdS_Ext_NH})
        \\ \cline{3-5}
        & & $L \to \pm \infty$ & Violation (Fig.~\ref{fig:DRN_AdS_Ext_LAM}) & Violation (Fig.~\ref{fig:DRN_AdS_Ext_LAM})
        \\ \cline{2-5}
        & \multirow{2}{*}{Non-Extremal} & $L = 0$ & \textbf{---} & $\alpha < 0$
        \\ \cline{3-5}
        & & \makecell{$L = 0$, \\ $\Lambda \to - \infty$} & \textbf{---} & $-2 < \alpha < - \frac{3 + \sqrt{17}}{2}$
        \\ \hline
    \end{tabular}
    \caption{Violations of the bound on chaos in the dilatonic Reissner--Nordstr\"om black hole background for the Einstein and string frames.}
    \label{tab:DRN}
\end{table}
\begin{table}[H]
    \centering \renewcommand{\arraystretch}{1.5}
    \begin{tabular}{|c|c|c|c|c|} \hline
    \multicolumn{3}{|c|}{\textbf{Kerr--Sen Black Hole}} & \textbf{Einstein Frame} & \textbf{String Frame}
    \\ \hline
    \multirow{7}{*}{$\mathbf{\Lambda = 0}$} & \multirow{5}{*}{Extremal} & \makecell{$L = 0$, \\ $r_0 = r_\mathrm{h} + \epsilon$} & \textbf{---} & \textbf{---}
    \\ \cline{3-5}
    & & \makecell{$r_0 = r_\mathrm{h} + \epsilon$, \\ $q = q^+$} & \makecell{$\frac{L}{m} < - \sqrt{a^2 + 2 a b}$, \\ $\sqrt{a^2 \! + \! 2 a b} \! < \! \frac{L}{m} \! < \! a \! + \! b$} & $\frac{L}{m} < - \frac{\sqrt{a^3 + 2 a^2 b + a b^2}}{a + 2 b}$
    \\ \cline{3-5}
    & & \makecell{$r_0 = r_\mathrm{h} + \epsilon$, \\ $q = q^-$} & $\frac{L^2}{m^2} > a^2 + 2 a b$ & $\frac{L^2}{m^2} > a^2$
    \\ \cline{3-5}
    & & $L \to \infty$ & \textbf{---} & \textbf{---}
    \\ \cline{3-5}
    & & $L \to - \infty$ & Violation & Violation
    \\ \cline{2-5}
    & \multirow{2}{*}{Non-Extremal} & $r_0 = r_\mathrm{h} + \epsilon$ & \textbf{---} & \textbf{---}
    \\ \cline{3-5}
    & & $L \to \pm \infty$ & $a > 0.7954$ (Fig.~\ref{fig:KS_LAM}) & $a > 0.7954$ (Fig.~\ref{fig:KS_LAM})
    \\ \cline{1-5}
    \multirow{4}{*}{$\mathbf{\Lambda < 0}$} & \multirow{2}{*}{Extremal} & \makecell{$L = 0$, \\ $r_0 = r_\mathrm{h} + \epsilon$} & $a > 0.9931$ (Fig.~\ref{fig:KS_AdS_Ext_NH}) & $a > 0.9944$ (Fig.~\ref{fig:KS_AdS_Ext_NH})
    \\ \cline{3-5}
    & & General Case & Violation (Fig.~\ref{fig:KS_AdS_Ext}) & Violation (Fig.~\ref{fig:KS_AdS_Ext})
    \\ \cline{2-5}
    & \multirow{2}{*}{Non-Extremal} & $\Lambda \to - \infty$ & \textbf{---} & \textbf{---}
    \\ \cline{3-5}
    & & General Case & $q > 9.9964$ (Fig.~\ref{fig:KS_AdS_NonExt}) & $q > 9.9042$ (Fig.~\ref{fig:KS_AdS_NonExt})
    \\ \hline
    \end{tabular}
    \caption{Violations of the bound on chaos in the Kerr--Sen black hole background for the Einstein and string frames.}
    \label{tab:KS}
\end{table}
    Tables~\ref{tab:DRN}~and~\ref{tab:KS} summarize the violations of the bound on chaos in the dilatonic RN and Kerr--Sen black hole backgrounds. For each frame, the tables show the specific parameter ranges in which violations occur. Cases in which violations are absent, or in which the bound on chaos cannot be evaluated, are denoted by dashes. The parameter independent violations are labeled as “Violation” in the tables, and cases examined numerically are referenced by the corresponding figures. These tables demonstrate the presence of frame dependence in the bound on chaos. Under specific conditions, violations are absent in the Einstein frame, whereas violations emerge in the string frame under the corresponding conditions. For large angular momentum, the dynamic contribution from the angular momentum dominates the dilaton contribution coupled to the particle mass, leading to identical values of the bound on chaos in both frames.

    Therefore, our analysis reveals the frame dependence of the bound on the Lyapunov exponent in dilatonic RN--AdS and Kerr--Sen--AdS black hole backgrounds. The results highlight the intrinsic sensitivity of chaotic behavior to the choice of conformal frame. Of particular interest are configurations in which the bound on chaos holds in the Einstein frame, while violations occur in the string frame, or vice versa. This frame-dependent behavior is expected to correspond to a distinct dynamical system and could provide insights into the underlying structure of holographic dualities. Our studies clarify the robustness of the bound on chaos, the role of frame dependence, and the interplay between black hole physics, string theory, and chaotic behavior. Future research is expected to address the persistence of bound on chaos violations under higher-order string corrections, the effects of particle backreaction, and the dynamics of extended string probes, including fundamental strings and D-branes.

\section*{Acknowledgments}
    This research was supported by Basic Science Research Program through the National Research Foundation of Korea (NRF) funded by the Ministry of Education (NRF-2022R1I1A2063176) and the Dongguk University Research Fund of 2025.

\printbibliography

\end{document}